# The threat of Centaurs for terrestrial planets and their orbital evolution as impactors

M.A. Galiazzo[1*] , E. A. Silber[2] , R. Dvorak[1]

[1]Department of Astrophysics, University of Vienna, Turkenschantzstrasse 17, 1180, Vienna, Austria

[2]Department of Environmental, Earth and Planetary Sciences, Brown University, 324 Brook St., Providence, RI, 02912, USA

**ABSTRACT**

Centaurs are solar system objects with orbits situated among the orbits of Jupiter and Neptune. Centaurs represent one of the sources of Near-Earth Objects. Thus, it is crucial to understand their orbital evolution which in some cases might end in collision with terrestrial planets and produce catastrophic events. We study the orbital evolution of the Centaurs toward the inner solar system, and estimate the number of close encounters and impacts with the terrestrial planets after the Late Heavy Bombardment assuming a steady state population of Centaurs. We also estimate the possible crater sizes. We compute the approximate amount of water released: on the Earth, which is about $10^{-5}$ the total water present now. We also found sub-regions of the Centaurs where the possible impactors originate from. While crater sizes could extend up to hundreds of kilometers in diameter given the presently known population of Centaurs the majority of the craters would be less than ∼ 10 km. For all the planets and an average impactor size of ∼12 km in diameter, the average impact frequency since the Late Heavy Bombardment is one every ∼ 1.9 Gyr for the Earth and 2.1 Gyr for Venus. For smaller bodies (e.g. > 1 km), the impact frequency is one every 14.4 Myr for the Earth, 13.1 Myr for Venus and, 46.3 for Mars, in the recent solar system. Only 53% of the Centaurs can enter into the terrestrial planet region and ∼ 7% can interact with terrestrial planets.

[*]Email: mattia.galiazzo@univie.ac.at

# 1.INTRODUCTION

Centaurs are objects in the solar system, with their orbits situated between those of the giant planets. There is the lack of general agreement among various definitions; thus, in this paper we will consider Centaurs as objects with a semi-major axis ranging from 5.5 au to 30 au. They are generally thought as a byproduct or an evolutionary product of the Trans-Neptunian objects, and thus should have generally lower density than the main belt bodies. This is because objects coming into the inner solar system from beyond Neptune can become comets more often than asteroids in the main belt and, as such should be consistent with volatile material. Centaurs have a short lifetime compared to the main belt asteroids, and are typically removed from the Solar System on timescales of only a few million of years (Di Sisto, Brunini & de Elía 2010; Galiazzo, Wiegert & Aljbaae 2016). These bodies are one of the main sources of Near-Earth Objects (NEOs). Thus, it is crucial to understand their orbital evolution, since it might end in a collision with one of the terrestrial planets and lead to catastrophic events. Some former comets which had performed many passages in the inner solar system could also have exhausted part of their water content and thus can be considered as hydrated asteroids, very likely of type C, D or P (Barucci et al. 1996). All this material can be delivered to the planets via impacts and become important for their atmospheres and have astrobiological implications. This work aims to : (i) constrain the rate of impacts on the terrestrial planets for the entire Centaur population in the recent solar system (after 3.8 Gyrs ago, Hartmann et al. 2012; Kirchoff et al. 2013), meaning after the Late Heavy Bombardment; (ii) find the typical orbits of the Centaur impactors; and (iii) determine from which sub-region of the Centaurs the impactors come from. Some past works have already attempted to compute this rate, but none of them considered the entire Centaurs population, e.g. Napier (2015); Napier et al. (2015). The paper is structured as follows: Section 2 describes the population sample, the tools to analyse the sample, and how the analysis was made. The results are given in Section 3: the source regions of the impactors are described in Section 3.1. Sections 3.2 and 3.3 show the statistical analysis and the results on close en-probability, interactions with the terrestrial planets, as well as a brief overview of interactions with the giant planets. In addition, we estimate the contribution of the Centaurs to the Near-Earth Objects (NEOs), Potential Hazardous Obcounters and the impact probability, interactions with the terrestrial planets, as well as a brief overview of interactions with the giant planets. In addition, we estimate the contribution of the Centaurs to the Near-Earth Objects (NEOs), Potential Hazardous Objects (PHOs) and Short Period Comets (SPCs). The conclusions are given in Section 4.

**2. MODEL**

We forward integrate 319 Centaurs (data[1] from JPL Small-Body Database Search Engine: http://ssd.jpl.nasa.gov/sbdbquery.cgi, see Table 1 for their osculating elements) for 30 Myr, which is a span of time larger than the average lifetime of Centaurs (typically removed from the solar system on timescales of only a few million years Dones, Levison & Duncan 1996; Levison & Duncan 1997; Tiscareno & Malhotra 2003; Horner, Evans & Bailey 2004a; Di Sisto & Brunini 2007; Bailey & Malhotra 2009; Galiazzo 2013), or until the body has undergoes an impact (see Table 2), or reaches a hyperbolic orbit and escapes. For each Centaur we compute 14 clones (the total of 4785 orbital evolutions, considering real initial orbits plus each clone orbits), distributed as described in Horner, Evans & Bailey (2004a); Galiazzo, Wiegert & Aljbaae (2016) where the osculating elements are: $a = a_0 \pm 0.005$, $e = e_0 \pm 0.003$ and $i = i_0 \pm 0.01$; here $a_0$, $e_0$ and $i_0$ are the initial elements.

**Table 1.** The range of osculating elements for the Centaurs sample considered in this work. The data are divided into regions (Region 1, outer region, and Region 2, inner region). Listed are (in this order) semi-major axis, eccentricity, inclination, perihelion and aphelion (minimum and maximum for the observed Centaurs in the range defined for the respective region).

| Region | a [au]        | e             | i                          | q [au]       | Q [au]        |
|--------|---------------|---------------|----------------------------|--------------|---------------|
| 1      | 13.15 - 29.92 | 0.009 - 0.942 | $2.0° - 170.7°$            | 1.42 - 27.38 | 15.67 - 56.43 |
| 2      | 5.50 - 13.02  | 0.022 - 0.867 | $2.0° - 175.2°$            | 1.37 - 12.31 | 6.37 - 22.99  |

For the Sun, we consider an impact when the body reaches a distance[2] of 0.00465 au relative to its center. Some comets were observed to survive in relatively close proximity to the Sun, such as C/2011 W3 (Lovejoy) which survived at a distance of 0.0055 au (Sekanina & Chodas 2012). We use the Lie-integrator (Hanslmeier & Dvorak 1984), an orbital numerical integrator with an adaptive step-size (Eggl & Dvorak 2010; Bancelin, Hestroffer & Thuillot 2012), also utilized in several previous works dealing with close encounters of asteroids with the planets (Galiazzo 2013; Galiazzo, Bazso & Dvorak 2014; Galiazzo & Schwarz 2014). In this work the accuracy parameter of the Lie-integrator is set to $10^{-13}$. The output stepsize for the numerical integration is 2 kyr, and a simplified solar system (SSS) is considered for the orbital propagation. In SSS all planets from Venus to Neptune are considered, with the mass of Mercury added to the mass of the Sun and assuming the Earth and the Moon are one body in their barycentric point[3]) Physical destruction and non-gravitational effects in the orbital propagation were neglected. We assume a close encounter (CE) occurs when a Centaur is within a distance from the center of the planet equal to $r_{CE}$ = 0.01 au from the perturbing body. This is because we aim to establish a region that can adequately cover the Hill sphere of the largest terrestrial planet (the Earth). However, for statistical analysis of the close encounters we separate terrestrial planets (TPs) from giant planets (GPs). For TPs, we consider the Hill radius of the Hill sphere as $r_{CE}$[4] 4 and, for GPs we scale to the largest hill sphere, the one of Neptune, taking into consideration the maximum value we have set for CEs, 0.01 au (see Table 2).

---

1 Updated at August 24, 2015.
2 695700 km, the volumetric mean radius of the Sun, see https://nssdc.gsfc.nasa.gov/planetary/factsheet/sunfact.html.
3 For the Hill sphere of the Earth, we considered the sum of the masses of the Earth and the Moon.

**Table 2.** Close encounter and impact radii per planet. ∗ means that the "Earth", is in reality the Earth and Moon together in the barycentric position.

| Planets | $r_{CE}$ [$10^{-3}$ au] | $r_{imp}$ |
|---|---|---|
| Venus | 6.71 | 0.04045 |
| Earth* | 9.88 | 0.04259 |
| Mars | 6.66 | 0.02266 |
| Jupiter | 4.38 | 0.46733 |
| Saturn | 5.33 | 0.38926 |
| Uranus | 5.84 | 0.16953 |
| Neptune | 10.00 | 0.16459 |

We consider the population comprised of observed Centaurs for the statistical computations and the probability results will pertain to a Centaur with the average size in the sample considered. Then, to mitigate bias, we only consider the unbiased (computed through a MonteCarlo simulation on the size distribution of the observed Centaurs Sheppard et al. 2000) population comprised of known Centaurs with D > 1 km. Based on the results from Sheppard et al. (2000), there are ∼ $10^7$ Centaurs, and, in the work of Horner, Evans & Bailey (2004a), km-size Centaurs should be 44300 (which later, see Section 3.1, will be the favourable value). We calculated the percentage of Centaurs that encounter the planets, $CE_{Ct,\%}$. The following was computed: the number of close encounters for all the Centaurs CE, the rate of encounters per million years, $CE_T$ [My$^{-1}$], the average time for close encounters with a planet, <T> [My], the time it takes for one respective body to experience a close encounter $T_{CECt}$ and the time required to undergo a close encounter per planet, $T_{CE}$ [My]:

$$CE_{Ct,\%} = N_{CE}/15/N_{T,CE} \cdot 100$$
$$CE = CE_C/15$$
$$CE_T = CE/T_{l,R} \quad (1)$$
$$T_{CECt} = (CE_{Ct,\%} N_{T,CE}/100/T_{l,R})^{-1}$$
$$T_{CE} = (CE/T_{l,R})^{-1}$$

---

4   A close encounter depends entirely on the relative gravitational attraction of each planet, and thus the close encounter region must be at least proportional to the relative Hill sphere; see also Galiazzo, Bazso & Dvorak (2013)

Here, $N_{CE}$, $N_{T,CE}$, CE C, $T_{l,R}$ are the total number of clones encountering the planet, the total number of observed Centaurs, the total number of close encounters for all the clones, and the average lifetime of a Centaur, respectively. Then, the following quantities were computed: the impact probability (see Section 3.3), and the average, maximum and minimum impact velocity for each TP, in order to esitmate the size of the craters, which are related to the impact velocity. The impact velocity ($v_{imp}$, Collins, Melosh & Marcus, 2005; Galiazzo, Dvorak, Bazso, 2014) was computed from the velocity at infinity (before the atmospheric entry, at the proximity of the Hill sphere border, $v_\infty$), taking into consideration the density, the diameter of the asteroid and the impact angle ($D_A$, $\rho_A$ and $\theta_{imp}$), the surface pressure and the gravitational acceleration of the given planet ($P_S$ and $g_P$):

$$v_{imp} = \frac{\rho_A D_A}{\frac{\rho_A D_A + 3 P_S}{2 g_P \sin \theta_{imp}}} v_\infty \qquad (2)$$

The mass of the body was obtained by assuming a spherical shape and the average density[5] 5 for Centaurs with a diameter less than 400 km: (940 kg/m$^3$). Taking into account the average impact angle of 45°(Le Feuvre & Wieczorek, 2008), we determined the impact velocity using the velocity at infinitum (as described earlier). These data, along with asteroid diameters, are needed to compute theminimum, average and maximum impact crater diameters on TPs. The asteroid diameter (see Table 2) was obtained from the Johnstone's archive (where in the case of multiple values, we selected the most recent one). If the diameter was not available from the archive, we computed it via the absolute magnitude and the average albedo (taken from Johnstone's archive) for Centaurs $\rho_V = 0.078$ using the standard equation of Tedesco et al., 1992):

$$D = \frac{1329}{\rho_V} 10^{\frac{-H_V}{5}}, \qquad (3)$$

where H is the absolute magnitude of the Centaur. The computations were done for 2 specific regions associated with Centaurs, and for the Centaurs population as a whole.

**Table 3.** Size of Centaurs per region in kilometers and relative masses in kilograms.

| Region | $D_{min}$<br>$M_{min}$ | $\bar{D}$<br>$\bar{M}$ | $D_{max}$<br>$M_{max}$ |
|---|---|---|---|
| 1 | 1.6<br>$1.769 \cdot 10^{13}$ | 51.5<br>$3.134 \cdot 10^{18}$ | 300.2<br>$1.066 \cdot 10^{20}$ |
| 2 | 0.5<br>$3.982 \cdot 10^{11}$ | 12.6<br>$7.396 \cdot 10^{17}$ | 72.0<br>$1.471 \cdot 10^{18}$ |

The two Centaur regions were divided based on the relative semi-major axis of the given minor bodies, because found by Galiazzo, Wiegert & Aljbaae (2016) there are 2 distinct average sizes borderlining J4:1 mean motion resonance (at 13.11 au). Beyond this resonance, the largest Centaurs (with the diameter larger than ∼ 100 km) are present in the outer part of the Centaurs' region. We define Region 1 (R1) as the region beyond J4:1 (13.11 au ≤ a ≤ 30 au), and Region 2 (R2) as 5.5 au ≤ a < 13.11 au.

---

5  We used the value given by the Johnstone's archive, http://www.johnstonsarchive.net/astro/tnodiam.html

Region 1 contains 160 known Centaurs and Region 2, 159. For the whole population, we summed up the probabilities and obtained the average of the average of the diameter in Regions 1 and 2. Thus, the probability is associated with a Centaur having that average value of diameter. We compute the amount of water released ($W_M$, in Earth masses per million years), assuming that the amount of water released by a single active Centaur, as a cometary body, is $P_C$ = 15% (this is consistent with the amount of water on the surface of the nucleus of comet 67P G-C, which is taken as a proxy for the whole body, Fulle et al. (2016), of its own mass), and $P_{NC}$ = 10% from other non-active Centaurs (asteroids), assuming their content to be similar to the most hydrated meteorite (Morbidelli et al., 2000). Guilberte-Lepoutre (2012) estimated that only ∼11% ($P_{CA}$, percentage of active comets) of the Centaurs would become active (comets). We used the average diameter (mass) per region and considering the rate of impacts from 0.1 Gyr ago until now and from the LHB until now, the probability was then multiplied by 1.49. Given that the population of Centaurs (such as their progenitors, the TNOs) from now until the LHB was on average 149% the present one and that in the last 0.1 Gyr it did not significantly change (only ∼1%)[6] ($I_r$ = 1/$K_r$ Myr, where $K_r$ is the impact rate in million years per each planet), it follows:

$$W_M = W_{M_C} + W_{M_{NC}}$$

$$W_{M_C} = \frac{4\pi \rho r_{c,region}^3}{3} I_r \cdot (P_C P_{C_A})$$

$$W_{M_{NC}} = \frac{4\pi \rho r_{c,region}^3}{3} I_r \cdot [P_{NC}(1 - P_{C_A})] \quad (4)$$

$$W_M = \frac{4\pi \rho r_{c,region}^3}{3} I_r \cdot [P_C P_{C_A} + P_{NC}(1 - P_{C_A})]$$

Here W M C and W M N C is the amount of water released per million year, respectively, by active and non-active Centaurs. Considering that we are interested in establishing (roughly) the amount of water released since the end of the LHB, this result was multiplied by the average number of bodies expected from that time until now. This results in 149% of the present population[7] (see Fig. 1 for the expected population from the LHB to the present). For the statistical computation of impacts, we also took into consideration the (average) amount of material lost per region and for the whole Centaurs population. Here we take into account the law proposed by Öpik (1976), and used also by Napier (2015), to determine the disruption (D kill ) in one dimension[8]:

$$D_{kill} = \frac{3.8}{\sqrt{\bar{q}}} \frac{\bar{T}_{life}}{\bar{T}_{Rev}} \, [\text{m}] \quad (5)$$

---

6  Data from Morbidelli from the work Brasser & Morbidelli (2013).
7  This is an extrapolation from the results for the TNOs population, which is the source region of the Centaurs (Brasser & Morbidelli, 2013; Galiazzo, Wiegert & Aljbaae, 2016)
8  For the purpose of our study this is sufficient, considering that this computation provides a rough estimate on which comet size might highly affect (completely or almost completely destroy) an active Centaur by solar radiation or tidal effect due to close encounters.

The expression above is a function of the perihelion (q̄) and apparition (perihelion passage). The latter represents the ratio between the average lifetime (<T`$_{life}$>) and the orbital revolution(<T$_{Rev}$>). We compute D kill when a given Centaur is supposed to be active (as a comet, or if it can become a comet) - this takes place when a Centaur remains inside the radius $a_c = a(1 − e^2) < 12$ au (see Guilbert-Lepoutre, 2012). This limit is appropriate because it is valid for Region 2, beyond Saturn (further away from the usual snowline, roughly ~ 5 au, as the model of Cyr, Sears & Lunine, 1998, asserts), and it has been determined to occur for real comets: the furthest active comet seen to-date is C/2017 $K_2$ (PANSTARRS), observed at a distance of ~ 24 au (Jewitt et al., 2017). We consider the average values (when $a_c < 12$ au) for all the clones of: (i) the average perihelion, q̄; (ii) the average period of revolution (from the averaged semi-major axis): <T$_{rev}$> = ā 1.5 y; (iii) the average lifetime, <T`$_{life}$>.

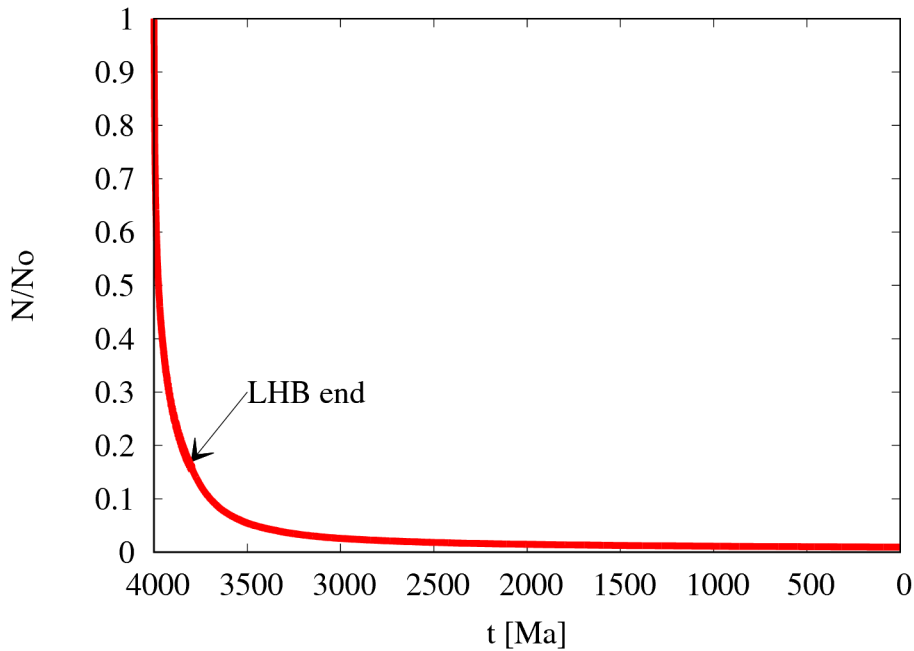

**Figure 1.** Decay of the Centaurs (from TNOs obtained in the model of Brasser & Morbidelli, 2013). This is a cumulative number of surviving minor bodies versus time in million years. The arrow shows the end of the LHB era, when the analysis of the Centaurs was initiated. The data comprising of the population decay of the TNOs was kindly provided by A. Morbidelli.

The impact crater dimensions can also be estimated using the analytical expression (see Collins, Melosh & Marcus, 2005):

$$D_{tc} = 1.161 \left(\frac{\rho_i}{\rho_t}\right)^{\frac{1}{3}} L^{0.78} v_i^{0.44} g_P^{-0.22} \sin^{\frac{1}{3}} \theta \qquad (6)$$

Here, $D_{tc}$ is the transient crater diameter, ρ i and ρ t are the impactor (projectile) and target density, respectively, L is the impactor diameter, $v_i$ is the impact velocity, $g_P$ is the gravitational acceleration for a given planet, and θ is the impact angle (here assumed to be 45 deg). The final diameter ($D_f$) for simple (bowl shaped) craters is given by: $D_f$ ~ $1.25 D_{tc}$ (Eq. (22) in Collins, Melosh & Marcus, 2005).

For complex craters, we use the expression that relates the transient crater diameter to the diameter associated with the simple-to-complex transition (D c ) on a planetary body (see Eq. (27) Collins, Melosh & Marcus, 2005):

$$D_{fr} = 1.17 \frac{D_{tr}^{1.13}}{D_C^{0.13}} \qquad (7)$$

For the sake of completeness, we also performed numerical simulations of impact crater formation using iSALE-2D, a multi-material, multi-rheology shock physics code (Melosh, Ryan & Asphaug, 1992; Ivanov, Deniem & Neukum, 1997; Collins, Melosh & Ivanov, 2004; Wünnemann, Collins & Melosh, 2006), an extension of the SALE hydrocode (Amsden, Ruppel & Hirt, 1980). To accommodate crater formation by impacts with a wide range of projectile sizes, in all our simulations we apply a two-layer model, with the crust and the mantle being represented with the equation of state (EOS) for granite (Pierazzo, Vickery & Melosh, 1997) and dunite (Benz, Cameron & Melosh, 1989), respectively. These are considered analogues for planetary crust and mantle (e.g., Yue et al., 2013). Since the projectile is considered to be of cometary origin, we applied the parameters appropriate for ice (see Bray et al., 2014, including the Tillotson EOS (Tillotson, 1962; Ivanov et al., 2002). Due to axial symmetry, all impacts are vertical, with the impact velocity corresponding to the vertical component of the velocity calculated earlier in this section. The crustal thickness ($d_{crust}$) and thermal gradient (dT /dz) for the three planets are as follows: Venus, $d_{crust}$ = 25 km, dT /dz = 10 K/km (Grimm & Solomon, 1988; James, ZUber & Phillips, 2013); Mars $d_{crust}$ = 50 km, dT /dz = 8 K/km (Zuber et al., 2000, Neumann et al., 2004); and the Earth d crust = 32 km, dT /dz = 25 K/km (e.g. Christensen & Mooney, 1995 and references therein). The grid resolution was varied according to the impact size, with the span of 10–30 cells per projectile radius (CPPR). Such choice in resolution is adequate for validating impact crater dimensions, while reducing computational expense. To prevent numerical instabilities, the high-resolution zone was set up such that it extends sufficiently far from the crater rim horizontally and below the transient crater floor vertically. The simulations were ran as long as it was required for the crater collapse to seize. In simulation time, this could be anywhere from a few hundred seconds up to several hours (basin size impacts). The crater diameter was measured from rim to rim, following the methodology described in Silber & Johnson (2017). Considering that the crater size obtained through simulations is consistent with that obtained using Equations 6 and 7, it was sufficient to preform numerical simulations for a representative sample of impacts of varying sizes, in order to capture a dynamic range of impact craters on the terrestrial planets.

## 3 RESULTS

From numerical orbital integrations, we find that Centaur clones undergo close encounters with all the planets; however, real impacts take place only with GPs, but not with TPs. Thus, to estimate the impact probability with TPs, we extrapolate it via fitting the number of close encounters per relative distance to the planet inside its Hill's sphere. We also compute the probability for Centaurs with D > 1 km (as mentioned in Section 2, these respresent the current population).

### 3.1 Source regions for terrestrial planets encounters in the Ceuntaurs' region

The entire Centaurs region is very chaotic, apart from some constrained orbital sub-regions. The source sites associated with terrestrial planets encounters (TPEs) can be seen in the brightest sections of Fig. 2.

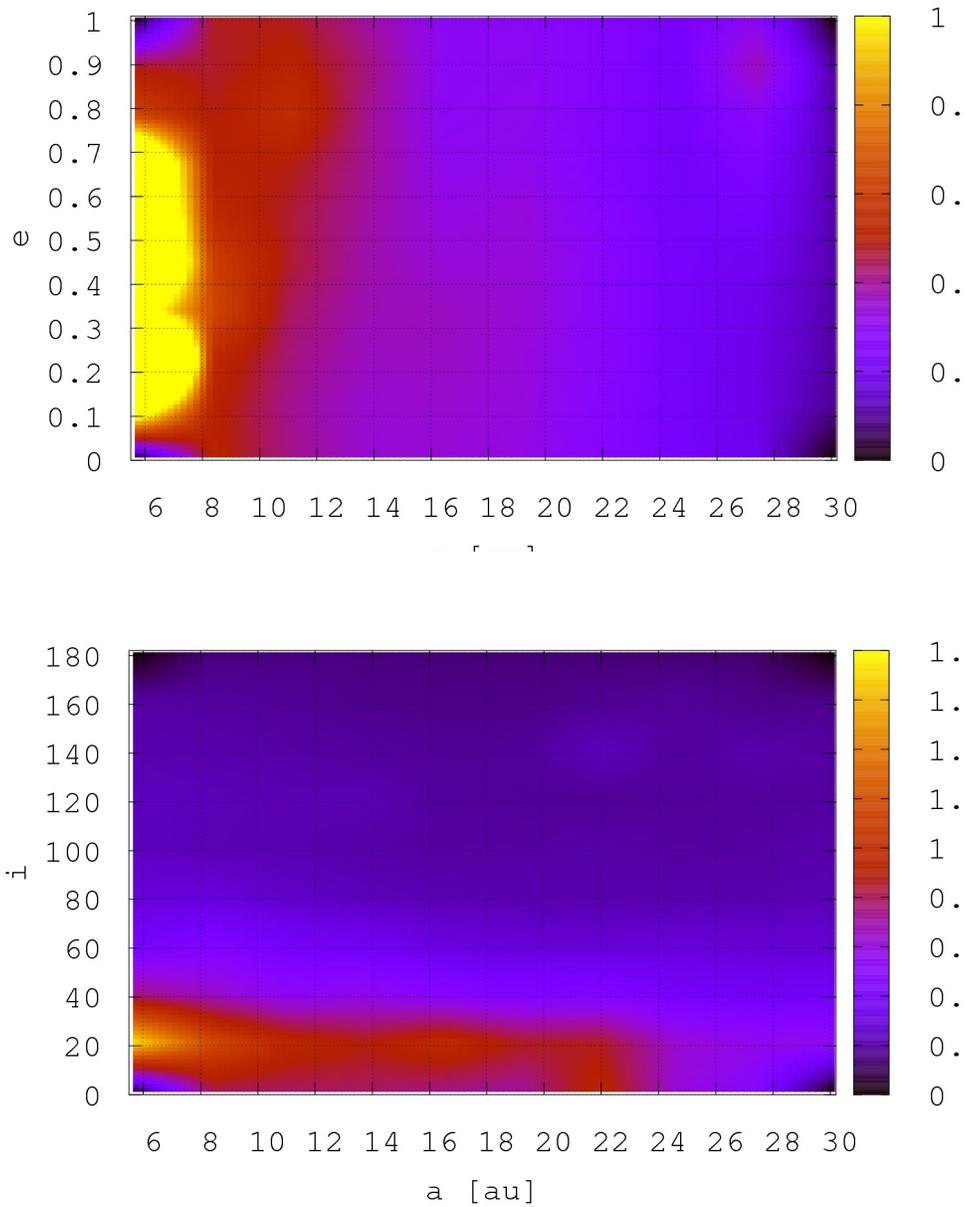

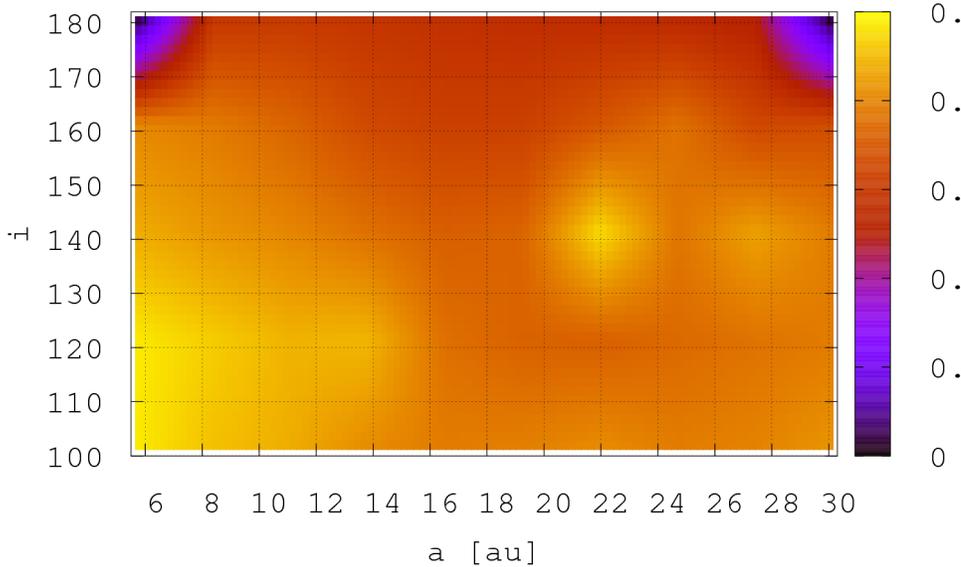

**Figure 2.** Source regions in the a − e and a − i space. The color code is related to the density of the relative points in each region, gaussian scaled. Note that the color bar scale is different in each panel. The full color figure is available in the on-line version of this paper.

The sub-regions which produce more TPEs are all of Region 2 (specifically, in the sub-region with semi-major axis, 5.5 au ≤ a ≤ 8 au and with eccentricity, 0.09 ≤ e ≤ 0.75. There is also a minor contribution from Region 1 in the following 2 phase space sub-regions: (i) a ≥ 20 au and e ≤ 0.63, (ii) 26 au ≤ a ≤ 28 au and 0.8 ≤ e ≤ 1.It is interesting to point out the inclination, where there is initially a predominant region for 8° ≤ i ≤ 40°, subsequently decreasing to a~24 au at i~30°. There is a minor contribution, although significant compared to other sub-regions in R1, in the sub-region for retrograde orbits at i ~140° at 22 au (see Fig. 3.1, bottom panel). However, the most predominant inclination is i~20°, similar to that of the Jupiter-family comets. Most of Centaurs are the sources of Jupiter family comets, (e.g., Fig. 3), where a Centaur becomes a Jupiter-family comet at 24 kyr (its Tisserand parameter in respect to Jupiter ($T_J$) is between 2 and 3, which by definition, corresponds to Jupiter family comets). Thus, all the sub-regions are also the major sources of possible impactors. The average lifetime for different regions varies. In some, the scattering from the average value is very large, and some also have very short lifespan (see i.e. Fig. 3), of the order of only thousands of years. We find the following lifetimes for Region 1, Region 2 and the whole region (all the Centaurs), respectively: $T_{life,1}$ = 10.158 Myr, $T_{life,2}$ = 1.824 Myr and $T_{life,TOT}$ ~ 6.041 Myr. This value is much higher than that found by (Napier, 2015), however they used only Chiron-like orbits and performed orbital evolutions for only 1 Myr (in our case, the average Chiron lifetime is 3.884 Myr) as a representation of the entire Centaurs population in fact Chiron is also in R1, a region where the Centaurs have an average larger life than the ones in R2. Of all the Centaurs population, 6.2 % collide with the Sun, 82.5 % escape the outer solar system in hyperbolic orbits, about 0.9 % have impacts with planets (Napier, 2015 computed 86%, in accordance with our result) and the rest survive 30 Myr. Region 2 is the region that experiences more impacts with the Sun, 56.4% of the total number of impacts with the Sun of the all Centaurs. This region also has relatively more impacts with GPs, and it dominates (as source of impactors again) for TPs (the latter one "statistically"), the ratio[9] is 1.6 and 20.7, respectively (see Section 3.3).

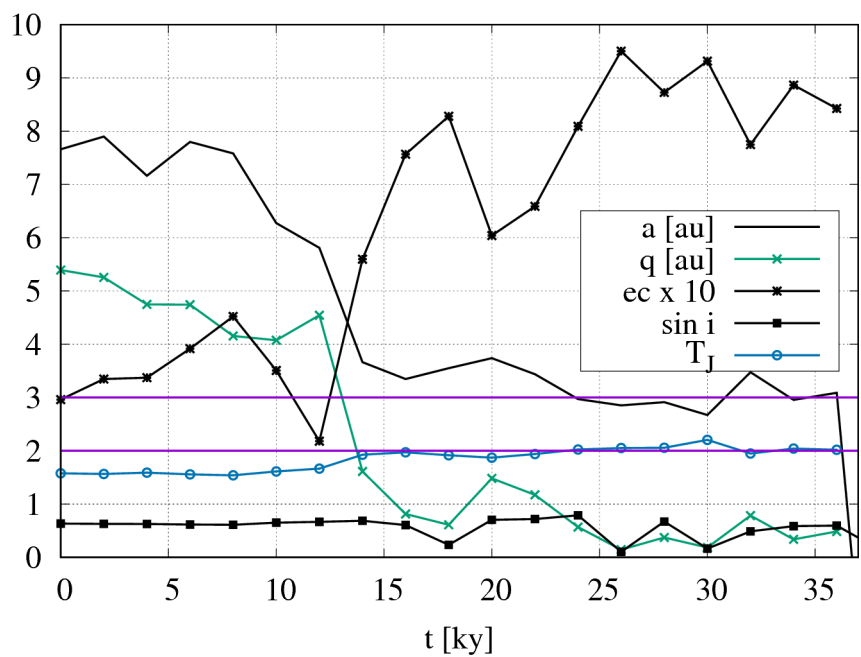

**Figure 3.** A short-life Centaur (coming from a source region of TPEs) which becomes a Jupiter family comet and then collides with the Sun. Plotted are the semi-major axis (a), perihelion (q), eccentricity (e), sinus of inclination (i) and Tisserand parameter in respect to the time. Horizontal lines at values 2 and 3 of y-axis represented the borders for Tisserand parameters ($T_J > 3$ for typical asteroidal orbits and $2 \leq T_J \leq 3$ for Jupiter family comet orbits).

**3.2 Close encounters**

CEs with Centaurs occur for all the planets, as shown in Table 4, where we can gauge that the Earth undergoes most encounters. We found that about 53% of the Centaurs enter the terrestrial planet region and of these, 57% are from Region 2 and the rest are from Region 1. The Centaurs, in general, arrive to the planets up to a couple million years (Table 4) comparable to that of the main belt asteroids (MBA) becoming NEOs (see Table III of Bottke et al., 2002). This demonstrates that the contribution of the Centaurs to the NEOs population is very important. consistent with a mixture of orbital evolutions of (former) MBAs and Centaurs. The latter usually evolve much faster as NEOs, only about 0.2 Myr, and consequently the Centaur-NEOs have shorter lifetimes. The whole population of NEOs survives for $T_N$ ~10 My, (Gladman, Michel & Froeschlé, 2000).

---

9   The ratio between the number of impacts per clone per region for the giant planets and the impact probabilities for terrestrial planets.

**Table 4.** Close encounter statistics for the clones in region 1, 2 and the whole population of Centaurs. "R1", "R2", "TOT and "D 1 " stand, respectively, forregion 1, region 2, the entire region (all the observed Centaurs) and all the Centaurs with D > 1 km (debiased population with 10 7 or 44300 and current rate: the average rate since the LHB has to be scaled by a factor of 1.49). $CE_{Ct\%}$, CE, $CE_T$ represent, respectively, the percentage of Centaurs (clones) having close encounters inside the Hill's sphere of the planet encountered, the total number of close encounters inside the planet relative Hill's sphere, the rate (over the lifetime of a Centaur in the region) of close encounters (inside the planet Hill's sphere) per time in million years, T is the average time of a close encounter with its minimum and maximum. $T_{CE\,Ct}$, $T_{CE}$ are the Centaur's rate arrival in a planet crossing like orbit and rate of close encounters in million years, respectively.

| Region/Planet | $CE_{Ct}\%$ | CE | $CE_T$ [$My^{-1}$] | T [My] | $T_{CE_{Ct}}$ [My] | $T_{CE}$ [My] |
|---|---|---|---|---|---|---|
| R1 Venus | 3.9 | ≈ 13 | ~ 1 | 4.470 (0.016-25.010) | ~ 1.655 | ~ 0.769 |
| R2 | 10.4 | ≈ 51 | ~ 28 | 1.644 (0.003-26.097) | ~ 0.115 | ~ 0.036 |
| $D_1$ | 7.2 | ~ 2022989  ~ 8962 | ~ 335043  ~ 1484 | ~ 3.057 | ~ $8.423 \cdot 10^{-6}$ | ~ $2.99 \cdot 10^{-6}$  ~ $6.74 \cdot 10^{-4}$ |
| R1 Earth | 4.3 | ≈ 13 | ~ 1 | 3.767 (0.023-24.942) | ~ 1.493 | ~ 0.801 |
| R2 | 10.7 | ≈ 42 | ~ 23 | 1.440 (0.004-26.201) | ~ 0.107 | ~ 0.043 |
| $D_1$ | 7.5 | ~ 1719958  ~ 7619 | ~ 284856  ~ 1262 | ~ 2.604 | ~ $8.070 \cdot 10^{-6}$ | ~ $3.51 \cdot 10^{-6}$  ~ $7.92 \cdot 10^{-4}$ |
| R1 Mars | 4.1 | ≈ 33 | ~ 1 | 3.555 (0.017-24.925) | ~ 1.586 | ~ 0.962 |
| R2 | 10.0 | ≈ 33 | ~ 18 | 1.395 (0.170-13.892) | ~ 0.113 | ~ 0.056 |
| $D_1$ | 7.1 | ~ 1343783  ~ 5953 | ~ 222554  ~ 986 | ~ 2.375 | ~ $8.548 \cdot 10^{-6}$ | ~ $4.49 \cdot 10^{-6}$  ~ $1.01 \cdot 10^{-3}$ |

The Centaurs are a relevant source of NEOs, considering that these objects have typical lifetimes  We also found that 43.4 % of the Centaurs become NEOs during their lifetime. According to the data available from HORIZON JPL (https://ssd.jpl.nasa.gov/sbdb_query.cgi, 977 NEOs with a diameter[10] larger than 1 km are known(and observed) now. It is thought that about 10% of the population still remains undiscovered (https://cneos.jpl.nasa.gov/stats/ and Tricarico 2017). From these values we can compute approximately the present contribution of Centaurs to km-size NEOs. We find that the observed population of Centaurs should produce 60 NEOs from Region 1 and 78 NEOs from Region 2, respectively, with the average dynamical time of 96 kyr for both (Table 5).

**Table 5.** Typical orbits and lifetime of Centaur-NEOs. Each second row per zone represents the minimum value for the respective osculating elements.

| Zone | $\bar{a}$ [au] | $\bar{e}$ | $\bar{i}$ | $\bar{T}$ [ky] |
|---|---|---|---|---|
| All | 2606.2 | 0.96 | 92.5 | 96 |
|  | 2.5 | 0.51 | 0.3 |  |
| ETR | 3.4 | 0.78 | 32.9 | ~15 |
|  | 2.5 | 0.51 | 0.3 |  |

Because the predicted numbers of NEOs from Bottke et al. (2002) are extrapolated, using a source zone with these borders: a < 4.2 au, e < 1.0, and i < 90° (a zone called ETR, Extended Target Region, from Bottke et al., 2002). We cannot compare our results with the standard debiased population of NEOs (because Centaurs were not considered). Therefore to compute the NEO contribution of Centaurs of kilometer size, we make a rough estimate, considering NEOs in the ETR region only. The number of (ETR-) NEOs known now is 239. As for Centaur-NEOs (CNEOs) found in the "Bottke's region", there are 20 from R2 and 11 from R1 with a dynamical life time of 21.3 kyr and 35.8 kyr respectively. Considering the entire lifetime of these objects, the frequencies of such episodes are 31/6.038 CNEOs per Myr, about 5 every Myr, or in terms of the average lifetime for the whole NEOs population, 50 CNEOs per 10 Myr. However, the probability to see them together with the other NEOs is low because the whole population of CNEOs in the ETR should live on average about 28.5 kyrs. Thus the debiased population present in the ETR region should be currently 50*44300* 0.0285/10 CNEOs[11] (we scale the probability to the average dynamical time to that of NEOs) and 239+239*0.01=253 (10% still unknown), the contribution of CNEOs to the current population of NEOs should be ~4%. This is consistent with Morbidelliet al. (2002), who assert that the contribution from the "comets" should be up to 10% of all the NEOs. However, a debiased estimate of NEOs for large eccentricities is also needed. As shown in Fig. 4, no NEOs are observed at e > 0.96 and beyond Jupiter very few. These results suggest that there should be more NEOs to be discovered than currently predicted. The unknown 1km size NEOs (10% of the current estimation) represents only a minimum. We found that the Centaurs have the highest probability of encountering the Earth, followed Venus and then Mars, if we countper single body. However, if we count the total number of close encounters, Venus has a slightly larger probability of close encounter. The probability of an Earth encounter for a body larger than 1 km is 7.5% (Table 4). The rate of arrival of large bodies (from Region 1, where the average size is D ∼ 50 km) in terrestrial planet region, is about 27.58 M yr$^{-1}$, which translates to 1 every 0.04 Myr. For the Earth-crossing orbits (close encounters inside the Hill sphere) this rate is 1.3M yr −1 (1 every ~0.8 Myr). Napier (2015), for > 100 km Centaurs, found 1.9 Chiron$^{-1}$Myr$^{-1}$ (1 every 0.5 Myr). However, that study considered any Centaur with a perihelion smaller than the semi-major axis of the Earth as an Earth-crosser. In our work, we found similar results (but more frequent), suggesting that Chiron-clones can have 185 Earth-crossing epochs in all, meaning that 185/15/3884 Chiron$^{-1}$ Myr$^{-1}$ or 1 every ~0.3 Myr.For R2, the rate of arrival is about 19 times greater than that for R1, with a similar rate for Venus (21) and Mars (17). There is a general trend of increasing rate with the decreasing distance to the Sun. In principle, Venus undergoes more close encounters than the Earth, and the Earth experiences more encounters than Mars (Table 4,), and each planet at least 1 every 50 kyr for the biased population. For Centaurs with D > 1 km, there is a close encounter approximately every 3-5 yr (3.5 yr for the Earth).Napier (2015) found that integrating the orbits of 100 Chiron clones, leads to the mean total dynamical lifetime on an Earth-crossing configuration of ~17.4 kyr, with a highly skewed distribution, and the median lifetime of 6.7 kyr. We found 3.767 Myr for Region 1, and 1.440 Myr for Region 2. This implies that Chiron should evolve relatively quickly in the planetary region and sooner than an average Centaur (for Chiron in particular, we found a mean total dynamical lifetime on an Earth-crossing configuration of ~1010 kyr, an order of magnitude greater than Napier, 2015). Fig. 5 shows the rate of close encounters of the clones of the known Centaurs every 1050 kyr. It is evident that in the first 3.5 million years (and in particular up to 1 Myr) the evolution is the most important factor for interactions with the Earth. After that, and for the next ~10 Myr Centaurs play an important role for R1. This implies that in a window of about 25 Myr, there is a possibility of high rate of close encounters, which is higher in 2 intervals of time at the beginning and at ~11 Myr from the initial orbital evolution.

---

10  If the diameter is not known, we selected the objects absolute magnitude H < 17.75.

---

11  We use the debiased value found by Horner et al. (2004), otherwise the number of CNEOs would be unreal, unrealistically larger than the present population of NEOs.

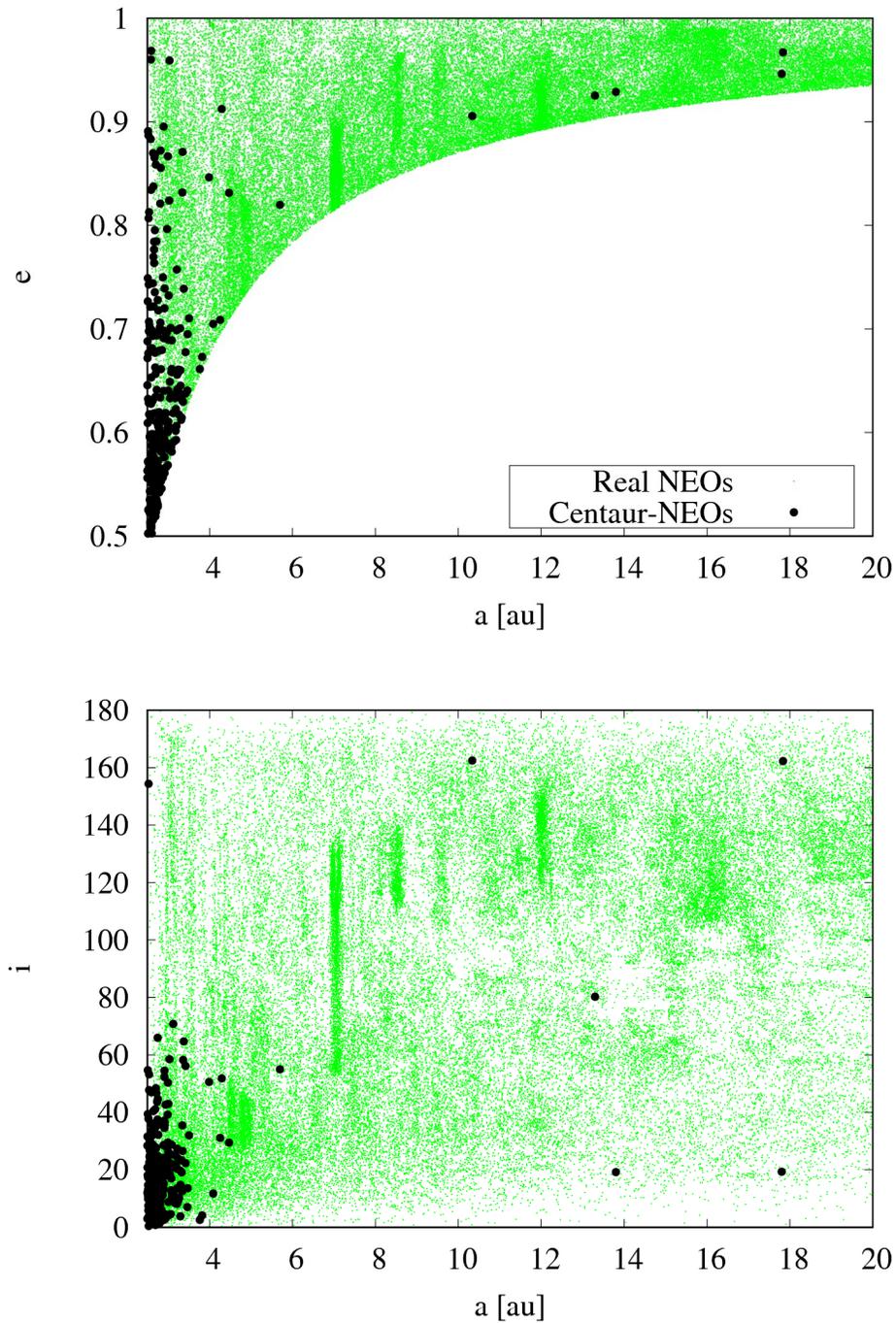

**Figure 4.** (a-e) and (a-i) phase space of real NEOs and all the possible orbits of the Centaur-NEOs (all clones).

Again, this is due to Centaurs from R1; the average time to enter the planetary region from R1 is much longer than R2, 4414 kyr and 722 kyr, respectively. The median values are 1418 kyr and 100 kyr, respectively, for R1 and R2, meaning that more than the half of large Centaurs from R1, arrive after the first peak.

The converse is also true for R2, where more than half of Centaurs entering the planetary region arrive in the interval time of the first bin (Fig. 5).

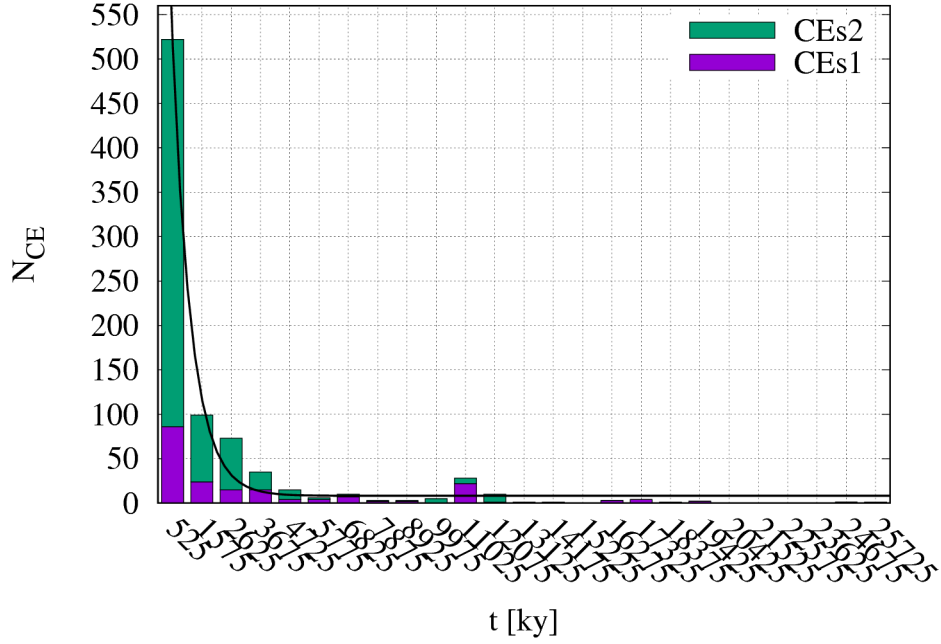

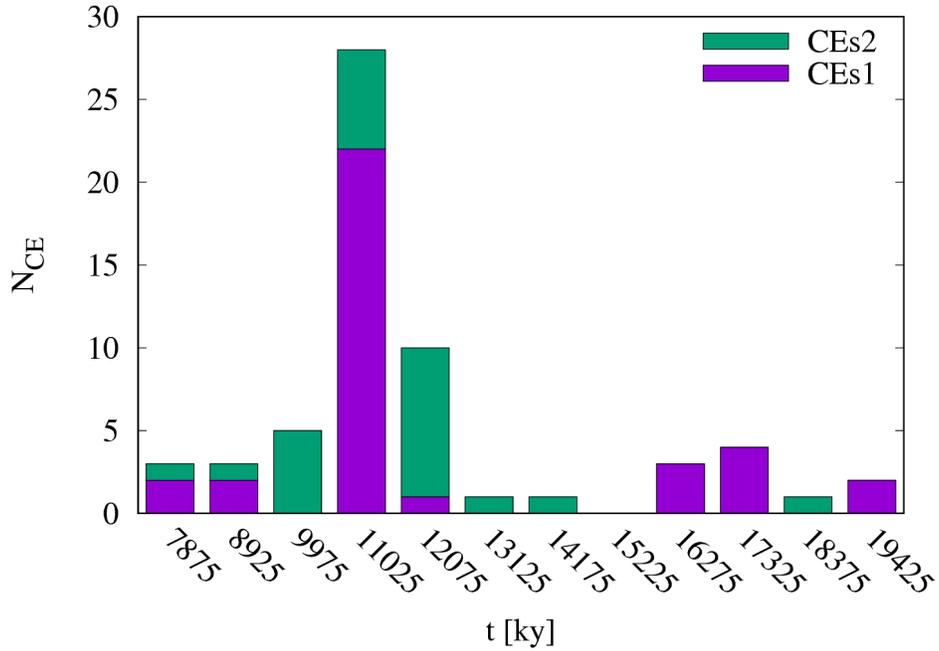

**Figure 5.** Rate of close encounters of the Centaurs clones with the Earth per kyr. Every bin (one every 1050 kyr) considered includes the contribution from the clones from the 2 different regions. Upper plot: the period of time when there are close encounters. Bottom plot: magnification of the time interval around the second peak.

An exponential decay can be seen in the entire population. We consider an exponential fitting function (the proportional rate growth):

$$N_{CE} = N_{CE,0} - \frac{a}{k}\left(1 - e^{kt}\right) \qquad (8)$$

and we found $N_{CE,0}$ = 1109.53 ± 77.13, a = 1.6086 ± 0.2385 and k = 0.001461 ± 0.000115. Some real (PHOs) are on very eccentric and large semi-major axis cometary orbits, thus they might belong to former comets, due to a break-up event (as also found in past studies, e.g. Horner, Evans & Bailey, 2004a, 2004b; Galiazzo, Wiegert & Aljbaae, 2016). In this work, we find those objects with a ≥ 2.6 au and e > 0.63 (see Fig. 6), but in particular with a > $a_J$ for Region 2 (i.e. asteroid 1999 XS 35 which could be a former comet or a piece of it coming from Region 1, considering that its size is small ~1 km, having an H=17.2). We also found 3 sets of Centaur Earth encounters with the following orbits (Table 6) in the "main belt orbital plane": (i) a ~ 2.7 au, e ~0.64 and i ~ 9°, i.e. asteroids 2016 $UF_{101}$ , 2008 VE, 2016 $YJ_4$, 2016 $EV_{27}$ and 2008 $DV_{22}$, all asteroids smaller than 100 m in diameter (H ≤ 21.8); (ii) a~ 3.2 au, e ~ 0.76 and i ~ 15˚. Asteroids closest to the average values are 2012 $DG_{61}$, 2008 $OB_9$. In terms of comets with similar orbits, those with the closest osculating elements are 3D/Biela, 15P/Finlay, 2007 $T_2$ (Kowalski), 222 P/LINEAR, 364P/PANSTARRS, P/2009 $WX_{51}$ (Catalina) and of 141P/Machholz 2 (and 2-A and 2-D); (iii) a ~ 3.4 au, e ~ 0.77 and i ~ 18˚. In practice, this range includes the range of set 2, but with a higher trend in semi-major axis. Some minor bodies close to this orbital range are (those that are more likely to come from Region 2), e.g., 2012 $GS_5$ (a small asteroid) and the comets 73P/Schwassmann-Wachmann group and 103P/Hartley 2.

**Table 6.** Statistics (average and standard deviation) of the orbital parameters for the clones in Region 1 and 2, which undergo close encounters with the terrestrial planets. "R1" and "R2" stand for Region 1 and Region 2, respectively. For the Centaur clones having the Tisserand parameter range in the given region during the close encounter Column CE is the number, $CE_{\%R}$ and $CE_{\%T}$ are the percentage in respect to the total and to the relative region.

| Region (Planet) | CE | CE%R | CE%T | $\bar{a} \pm \sigma_a$ | $\bar{e} \pm \sigma_e$ | $\bar{i} \pm \sigma_i$ |
|---|---|---|---|---|---|---|
| R1 Venus $T_J > 3$ | - | - | - | - | - | - |
| $2 < T_J < 3$ | 63 | 32 | 7 | 3.16 ± 0.40 | 0.820 ± 0.068 | 17.4 ± 18.1 |
| $T_J < 2$ | 137 | 69 | 14 | 13.07 ± 19.89 | 0.92 ± 0.07 | 102.6 ± 55.7 |
| R2 $T_J > 3$ | - | - | - | - | - | - |
| $2 < T_J < 3$ | 185 | 24 | 19 | 3.22 ± 0.46 | 0.82 ± 0.06 | 18.7 ± 16.0 |
| $T_J < 2$ | 583 | 76 | 60 | 9.80 ± 14.10 | 0.91 ± 0.06 | 114.1 ± 52.0 |
| R1 Earth $T_J > 3$ | 6 | 3 | 1 | 2.70 ± 0.03 | 0.64 ± 0.01 | 8.98 ± 3.10 |
| $2 < T_J < 3$ | 92 | 48 | 11 | 3.23 ± 0.51 | 0.76 ± 0.09 | 14.6 ± 11.4 |
| $T_J < 2$ | 94 | 49 | 12 | 15.27 ± 19.05 | 0.93 ± 0.06 | 104.9 ± 54.7 |
| R2 $T_J > 3$ | - | - | - | - | - | - |
| $2 < T_J < 3$ | 176 | 28 | 22 | 3.42 ± 0.63 | 0.77 ± 0.09 | 17.5 ± 14.5 |
| $T_J < 2$ | 455 | 72 | 56 | 10.19 ± 12.05 | 0.89 ± 0.08 | 116.6 ± 52.0 |
| R1 Mars $T_J > 3$ | - | - | - | - | - | - |
| $2 < T_J < 3$ | 63 | 41 | 10 | 3.66 ± 0.78 | 0.69 ± 0.11 | 13.0 ± 11.6 |
| $T_J < 2$ | 91 | 59 | 14 | 18.73 ± 28.55 | 0.88 ± 0.09 | 111.8 ± 50.9 |
| R2 $T_J > 3$ | 9 | 2 | 1 | 2.91 ± 0.16 | 0.53 ± 0.07 | 10.6 ± 3.8 |
| $2 < T_J < 3$ | 155 | 32 | 24 | 3.98 ± 1.03 | 0.72 ± 0.11 | 19.3 ± 14.8 |
| $T_J < 2$ | 325 | 66 | 51 | 10.80 ± 9.31 | 0.85 ± 0.09 | 117.0 ± 49.9 |

The asteroids with similar orbits to set 1 and 2 might have originated from former comets coming from Region 1. Additionally, comets in set 2 come from Region 1. For set 3, there are bodies coming from Region 2. Typically, the Centaurs coming from Region 1 are on the edge of the Earth-crossing bodies in the a − e phase space, and they match the orbit of the Near-Earth-Comets (Fig. 6).

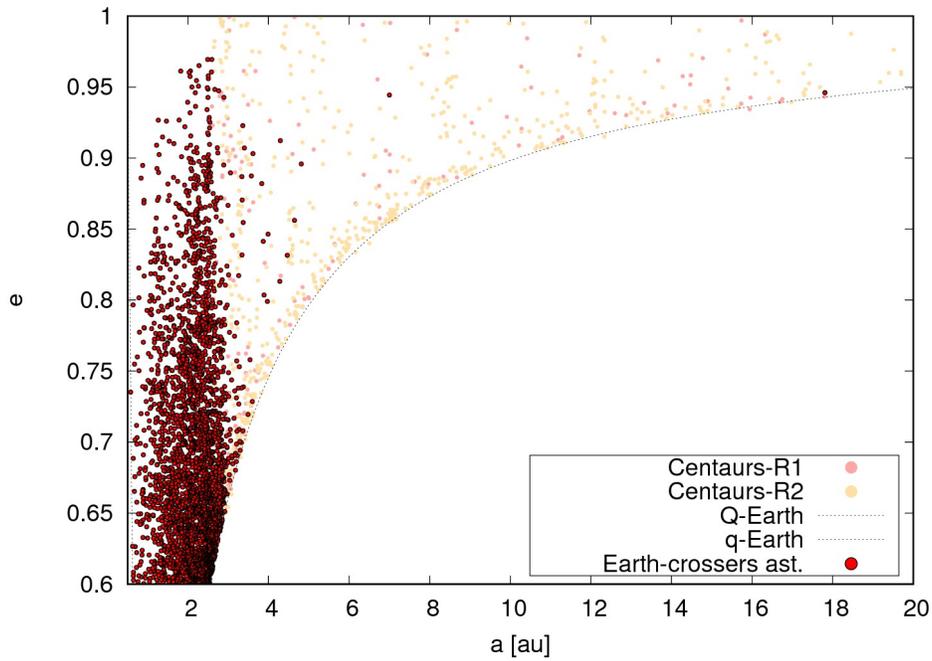

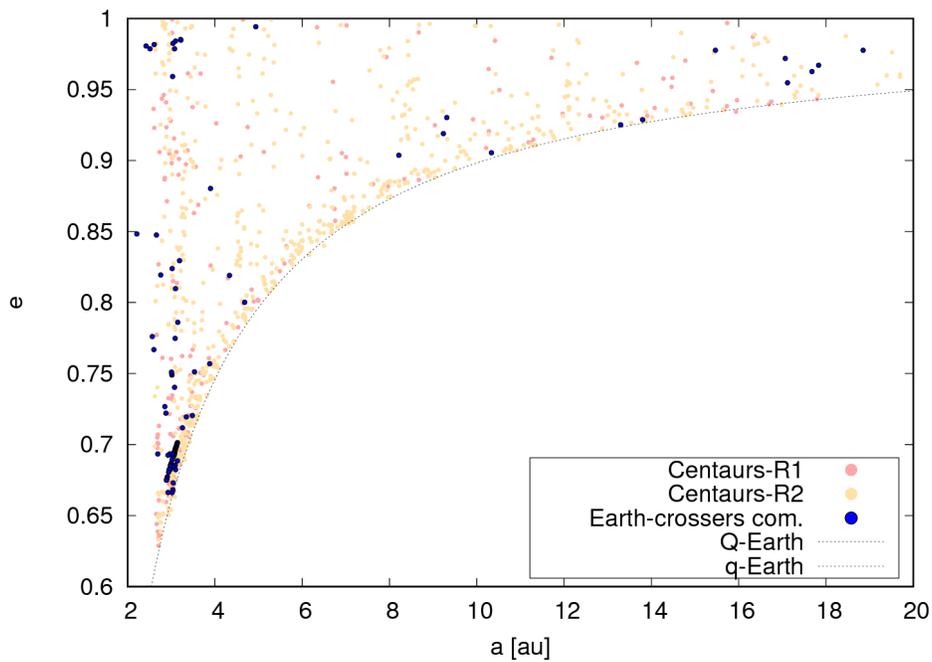

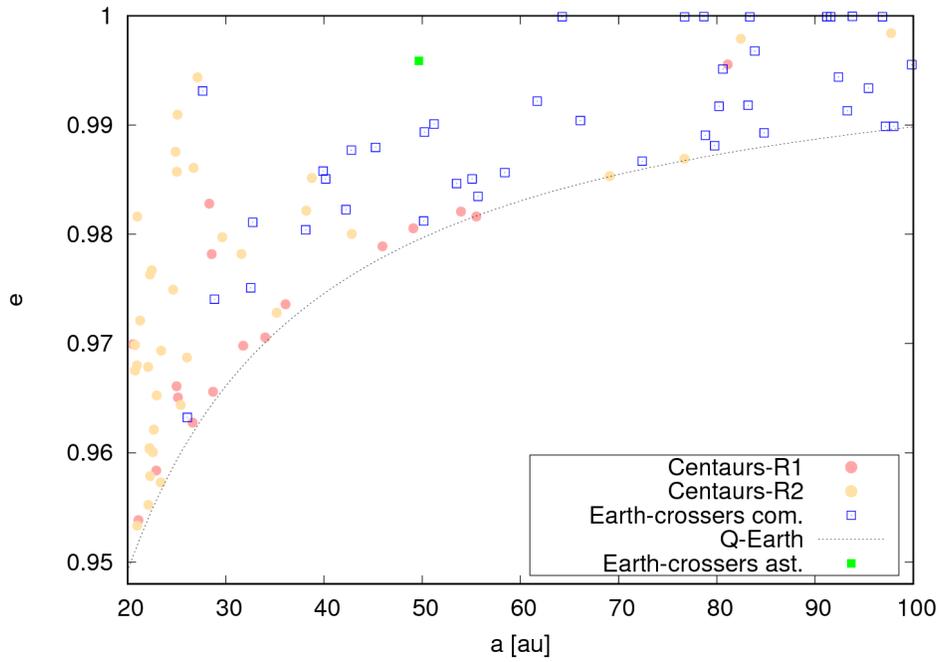

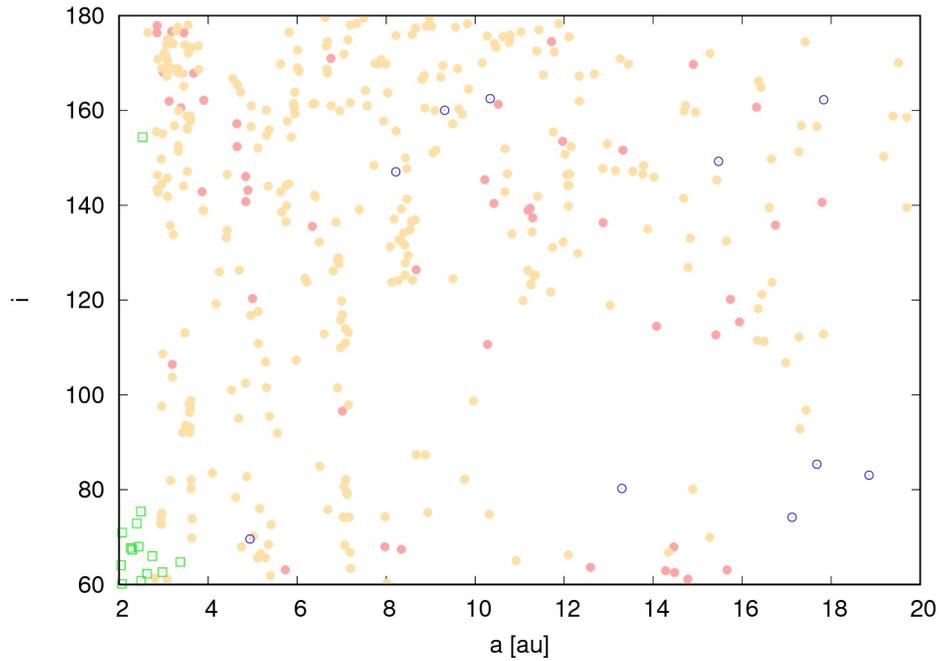

**Figure 6.** Centaurs' Earth encounters comparison with real NEOs which can cross the Earth's orbit in the a − e and a − i phase space. Centaurs-R1 and -R2 represent the orbit of the clones found in the regions in the plots. The Earth-crosser asteroids (Earth-crossers ast.) and comets (Earth-crossers com.) are the orbits of the real observed population. In the panel which describes the a − i phase space, the dotted square and the dotted circles represent the real observed asteroids and the real observed comets, respectively. Clones of Centaurs are also shown as in the other panels.

In the a-i space there are also preferred regions at certain values. It is interesting to point out a gap beyond $11^{+2}_{-1}$ au, in particular, at a specific value in retrograde orbits with the following inclinations range $85° \leq i \leq 110$. Unlike our clones, no currently known real NEO is seen at this range, corroborating these results. However, further studies, including observations, are needed to confirm the presence of this gap. From the numerical orbital evolution standpoint, it might be possible to cover this region by a larger number of clones. From the observational standpoint, this might be a bias due to difficulties of finding asteroids in this region. Nevertheless, we cannot exclude the possibility that this could be a real selection effect due to peculiar gravitational effect or due to the orbital evolution of the asteroids arriving in particular orbits into the planetary region. Integrations for regions at semi-major axis larger than 20 au do not show Centaur Earth encounters at low inclinations (below 70 degrees) like the real ones. This could be a bias deflection due to the integrator or an observational gap. We also note that on average, Centaurs coming from Region 1, thus the largest ones, remain at higher inclinations, see Fig. 6 and Table 6. The leakage from the main belt asteroids appears inadequate for NEOs larger than 3 - 4 km and it is also an order of magnitude less than that cited in the previous studies (Menichella, Paolicchi & Farinella, 1996; Minton & Malhotra, 2010). The Centaurs complete this lack of NEOs in this aformentioned size range. Thus the impact rate of the bodies of this size is complete with the contribution of the Centaurs added to the NEOs coming only from the main belt. From Table 6, we also found that the contribution to the (Potential Hazardous Objects) PHOs is at least of 1 object per Myr, and, applying the debiased model of Horner, Evans & Bailey (2004a), there should be at least 139 (= 44300/318) km-size objects per Myr. However the probability to see these objects is very low, considering that their dynamical time as NEOs, determined earlier, is 25 kyr. Thus, performing the same computation as done for all the NEOs population, leads to only one former Centaur or none, among the observed PHOs with a MOID[12] < 0.01 au (we consider the 11 observed PHOs inside 0.01 au because our numerical integration was limited to this planetocentric distance). In agreement of this finding, among the observed PHOs, there is only one comet, C/1983 $H_1$ (IRAS-Araki-Alcock), which has also very similar orbits (a = 98.02, e = 0.9899 and i = 73.25) to our computed CNEOs, see Fig. 6. Following the same approach, and considering that only ~11% of the Centaurs can become active, we can compute the contribution to the short periods comets, or objects with a period less than 200 years and a C < 19 au. We find that from R2 and R1, 2335 and 1743 clones, respectively, have a c in the "activity range" with a dynamical life time of ~581 kyr (R1) and ~406 kyr (R2). Given that we do not have an estimate of a unbiased population , we can at least compare the observed bodies: from our estimate, 30 (= 2335/158 + 1743/160 ∗ 0.11) short period comets should be produced by the observed population of Centaurs. From the Horizon JPL catalogue, we find 96 short period km-size comets, thus considering that the number of comets missing to get the unbiased value is larger for the Centaurs than the SPC, the contribution of the Centaurs to the SPC is ≥ 31%. These minor bodies arrive on average at cometary speeds (Table 7), which are generally higher than those for Centaurs coming from Region 1, due to their more eccentric orbits (Table 6).

---

12 Minimum Orbit Intersection Distance.

**Table 7.** Velocity at infinitum for a close encounter with each planet in kilometers per second. * and ** mean Region 1 and Region 2, respectively.

| Planet | $v_{min}$ | $\bar{v}$ | $v_{max}$ |
|---|---|---|---|
| Venus* | 12.4 | 48.9 | 84.8 |
| Earth* | 9.4 | 38.8 | 71.7 |
| Mars* | 7.2 | 30.7 | 59.1 |
| Venus** | 11.2 | 46.3 | 84.2 |
| Earth** | 9.8 | 40.9 | 71.5 |
| Mars** | 5.9 | 27.6 | 60.3 |

### 3.3 Impacts

Before computing the impact probability, we calculate the disruption rate for our sample using the Öpik relation Öpik (1976) with which we can find the new average diameter per region, considering only the active comets (11%). With this calculation, ~ 62.5% are in Region 1 (10 out of 16 given the results from Guilbert-Lepoutre (2012), their Table 3). We use the averaged version of the Öpik's relation:

$$r_{lost} = 3.8/\sqrt{q}/T_{rev} \quad [cm] \qquad (9)$$

Using the average perihelion (q) and orbital revolution period around the sun ($T_{rev}$) of the Centaurs which enter in the terrestrial region, r lost is the radius which loses the Centaur through its activity as a comet. The average period of revolution is <T> = $\bar{a}^{1.5}$, where a is the semi-major axis (in au taken through all its evolution, from the beginning of the integration to the end of its dynamical life) of the Centaurs which enter the terrestrial region. We determined that the average dismantle per planet in Region 1, Region 2 and the total region is overall negligible[13], having a maximum average disruption of about 200 m in diameter (Table 8).

**Table 8.** Average semi-major axis ($\bar{a}$), perihelion ($\bar{q}$), orbital revolution period (<$T_{rev}$>), the lifetime of a body when it is in a cometary orbit (when it can be active) $a_c$ < 12 au (<$T_{Opik}$>). All values are computed per relative region. These values provide the remaining diameter of the clones, where we report the values of the average disruption in one dimension <$D_{kill}$> and the average value found for diameters after disruption, <$D_O$>.

| Planet | $\bar{q}$ [au] | $\bar{a}$ [au] | $\bar{T}_{rev}$ [Myr] | $\bar{T}_{Opik}$ | $\bar{D}_{kill}$ [km] |
|---|---|---|---|---|---|
| R1 | 3.66 | 784.55 | 21975 | 724 | 0.065 |
| R2 | 3.79 | 325.16 | 5863 | 677 | 0.225 |

This result suggests that the surviving bodies of the known sample coming into the inner solar system are generally not affected by the activity: the main reason for disruption by the activity before encountering a TP could be due for some isolated cases, such as comets with a small nucleus (D ≤ 1 km), thus Centaurs with D << 1 km are not taken into account for statistical computations. Table 9 gives a summary of the impact probability per million years, impact velocity and diameter of possible craters.

---
[13] We consider the average "Öpik" disruption, and use the average parameters to compute it.

**Table 9.** Impact probability (IP) per million years and impact rate (K) in Myr for each planet: * means a period from now to the LHB, without * means from now until 0.1 Gyr ago. "I", "II" (for the observed population and thus for a body of their average diameter) and "C" stand for, respectively, Region 1, Region 2, the entire region (all the Centaurs) for the entire real (unbiased with with $10^7$ bodies ("S") and with 44300 ("H") population with a diameter larger than 1 km. Out means more than the solar system age (no possible impacts after the LHB). $W_{Earth}$ is the water release in Earth masses from the LHB until now. For the water released by the unbiased population, we considered an asteroid with D = 5 km, which is the mode value of all the diameters.

| Planets | $IP_I$ [$10^{-8} ky^{-1}$]<br>$K_I$ [My] | $IP_{II}$ [$10^{-8} ky^{-1}$]<br>$K_{II}$ [My] | $IP_C$ [$10^{-2} ky^{-1}$] - S<br>$K_C$ [ky] - S<br>W [$W_{C,Earth} 10^{-6}$] - S | $IP_C$ [$10^{-5} ky^{-1}$] - H<br>$K_C$ [ky] - H<br>W [$W_{C,Earth} 10^{-6}$] - H |
|---|---|---|---|---|
| Venus | 2.67 ± 0.01  3.98* ± 0.02<br>out  out* | 47.26 ± 0.01  70.42* ± 0.01<br>2116 ± 2  1420* ± 1 | 1.6 ± 0.1  2.3* ± 0.1<br>64 ± 1  43* ± 1<br>∼ 3.40  ∼ 19.2* | 7.0 ± 0.1  10.3* ± 0.1<br>out  965* ± 1<br>∼ 0.02  ∼ 0.9* |
| Earth | 2.26 ± 0.01  3.37* ± 0.01<br>out  out* | 52.69 ± 0.01  78.5* ± 0.01<br>1898 ± 1  1274* ± 1 | 1.7 ± 0.1  2.6* ± 0.1<br>58 ± 1  39* ± 1<br>∼ 3.7  ∼ 21.2* | 7.7 ± 0.1  11.4* ± 0.1<br>out  877* ± 1<br>∼ 0.02  ∼ 0.9* |
| Mars | 0.62 ± 0.02  0.09* ± 0.01<br>out  out* | 14.90 ± 0.01  22.19* ± 0.01<br>out  out* | 0.5 ± 0.1  0.7* ± 0.1<br>205 ± 1  138* ± 1<br>∼ 1.1  ∼ 6.0* | 2.2 ± 0.1  3.2* ± 0.1<br>out  3104* ± 1<br>∼ 0.01  ∼ 0.3* |
| Jupiter | | | 1112 ± 45  1657* ± 67<br>0.090 ± 0.002  0.060* ± 0.002<br>∼ 2406.4  ∼ 136247.6* | 4.9 ± 0.2  7.3* ± 0.2<br>20.03 ± 0.42  13.63* ± 0.28<br>∼ 10.7  ∼ 603.5* |
| Saturn | | | 278 ± 11  414* ± 17<br>0.360 ± 0.009  0.242* ± 0.006<br>∼ 601.6  ∼ 34061.9* | 1.2 ± 0.1  1.8* ± 0.1<br>81.21 ± 1.70  54.51* ± 1.14<br>∼ 2.7  ∼ 150.9* |
| Uranus | | | - | |
| Neptune | | | 35 ± 1  52* ± 2<br>2.878 ± 0.071  1.932* ± 0.047<br>∼ 75.2  ∼ 4257.7* | 0.2 ± 0.1  0.2* ± 0.1<br>64.97 ± 13.56  436.05* ± 9.34<br>∼ 0.3  ∼ 18.9* |

The probability of impacts for TPs was found by correlating the number of close encounters to the relative distance for all the binning values available (see Fig. 7), and fitting this data by a parabolic equation:

$$y = ax^2 \quad (10)$$

Here, y is the number of close encounters per given distance (x), and a is the relative fitting parameter. This parameter was subsequently adapted to the number of clones ($N_C$ =15) per Centaur and computed per million years (considering the average lifetime <$T_{life}$> of the Centaurs), such that the impact rate per million year is:

$$P = \frac{y}{15 \cdot \bar{T}_{life}}$$
$$P = \frac{ax^2}{15 \cdot \bar{T}_{life}} \quad (11)$$

The error associated with the measurement was computed with the partial derivatives:

$$\Delta P = \sqrt{\left(\frac{\partial P}{\partial a}\sigma_a\right)^2 + \left(\frac{\partial P}{\partial T_{life}}\sigma_{T_{life}}\right)^2}$$

$$= \frac{1}{15}\sqrt{\left(\frac{x^2}{T_{life}}\sigma_y\right)^2 + \left(\frac{ax^2}{T_{life}^2}\sigma_{T_{life}}\right)^2} \qquad (12)$$

$$\sigma_y = \frac{\partial y}{\partial a}\sigma_a = x_0^2 \sigma_a$$

The error for the average lifetime and for the fitting constant, a, are their standard deviation of the mean. $x_0$ is the radius of the planet engaged. Thus, the final probability of impacts per million years is $P = a * x^2_{imp}/15/\langle T_{life}\rangle \pm \Delta P\ [M\ y^{-1}]$.

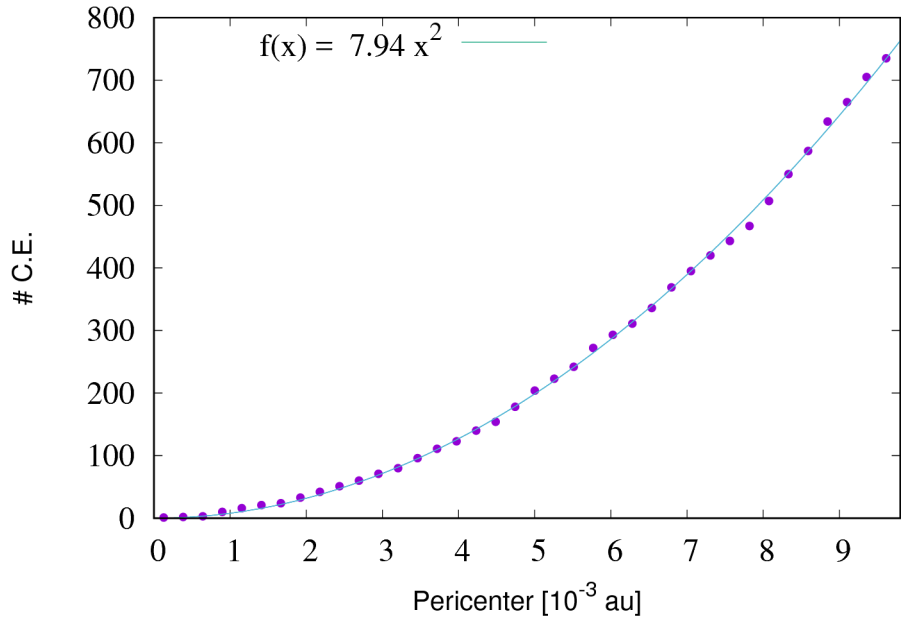

**Figure 7.** Fit of the data for close encounters of Centaurs from Region 2 with the Earth in its relative Hill sphere.

For GPs, we only count the number of real impacts (found innumerical orbital integrations) per given time, because we are more interested in TPs in this work. Table 9 shows the forecasted number of impacts per region of the terrestrial planets, the frequency and recurrence time expected per regions and global region (forCentaurs with D > 1 km). Centaurs from Region 1 do not seem to present any threat for any of the 3 TPs in the recent solar system (computations are for an impact in more than 3.8 Gyr). We find that no impact should happen from bodies from Region 1. In Region 1, even though there are bodies on average ~51 km in size, the possibility of an impact is even more unlikely. For example, Napier (2015) found that a 100 km comet striking the Earth would release ~1000x the energy required to form the 150 km Chicxulub crater, and would presumably remove the entire biosphere. The impact crater statistics show this scenario to had been an unlikely event over the 600 Myr time-scale of Phanerozoic, and for other results even longer than ~4 Gyr, assuming the current orbits of the presently known[14] planets in our solar system. As for Region 2, for an average Centaur of size 12.1 km, an impact on the Earth and Venus happen every 1.892 Gyr and 2.110 Gyr, respectively. This implies that 1 or 2 catastrophic impacts on these planets after the LHB could have been possible. These 2 impacts should create craters of size 133 km on Venus and 119 km on the Earth. Although it is not feasible to confirm whether such an impact had occurred on the Earth, it is interesting to entertain the possibility. Of all confirmed impact structures on the Earth, the Sudbury crater falls within the size range (age 1850 ± 3 Ma and D ~ 130 km). The Vredefort crater might have been created by a Centaur (Johnson & Bowling, 2014); however, if this is correct, then it follows that the Earth is right now within the interval of time susceptible to the collision with a large impactor. Thus, an extinction event could hypothetically occur at present and in a time window of about 60 kyr. A large crater like that would not come only from the supposed E-belt (Bottke et al., 2012) objects, but also from Centaurs. Additionally, a few of the Centaurs seem to become active. Out of nearly 150 Centaurs detected to date, the comet-like activity has been reported in only 16 of them, with the current fraction of active Centaurs being $f \sim 10^{-1}$. Therefore, many of Centaurs could potentially preserve most of their mass if they do not approach too close to the Sun. The rate of impacts for all the Centaurs having a smaller average size (roughly ~5 km, the mode value of the presently known Centaurs), Venus, Earth and Mars experience an impact in a time less than 64 kyr, 58 Kyr and 205 kyr, respectively. Thus, there is at least one impact in this limit. For an impactor of 1 km in size, the resulting craters on the Earth would range from 6.9 km (see Fig. 8) to 18.9 km in diameter. The transport of water is negligible compared to the total present amount of water (~ $5*10^{-4} M_{Earth}$, Lecuyer, Gillet & Robert, 1998; Morbidelli et al., 2000). For the Earth, it should be on the order of at least $10^{-4}$ times the current amount of water, which implies that the water present now comes mostly from other sources or minor bodies older than 3.8 Gyr.

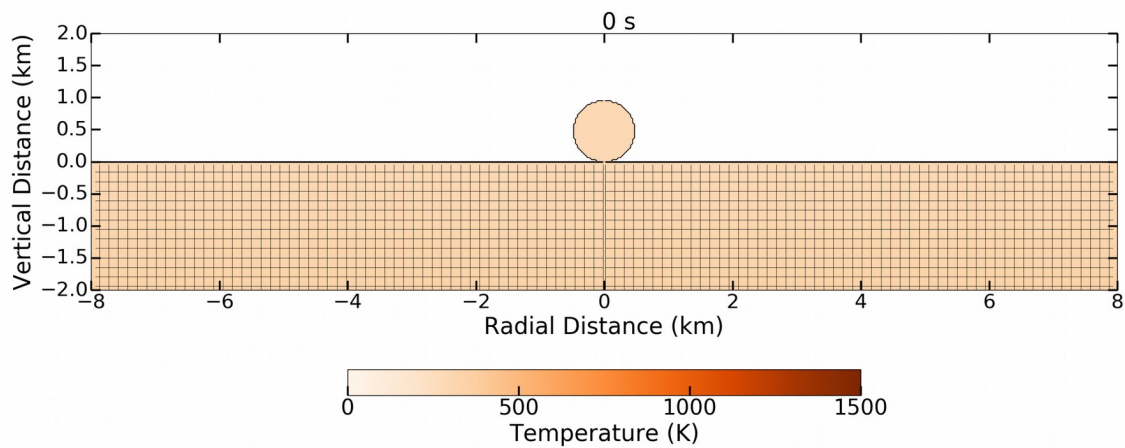

---

14  Escluding, i.e. a possible, planet IX, or more after Neptune (Gomes, Matese, & Lissauer, 2016).

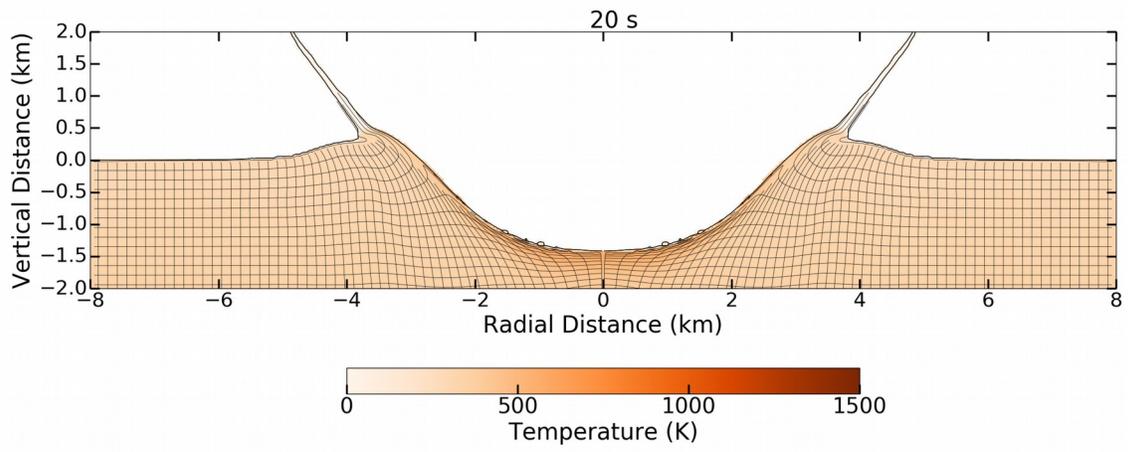

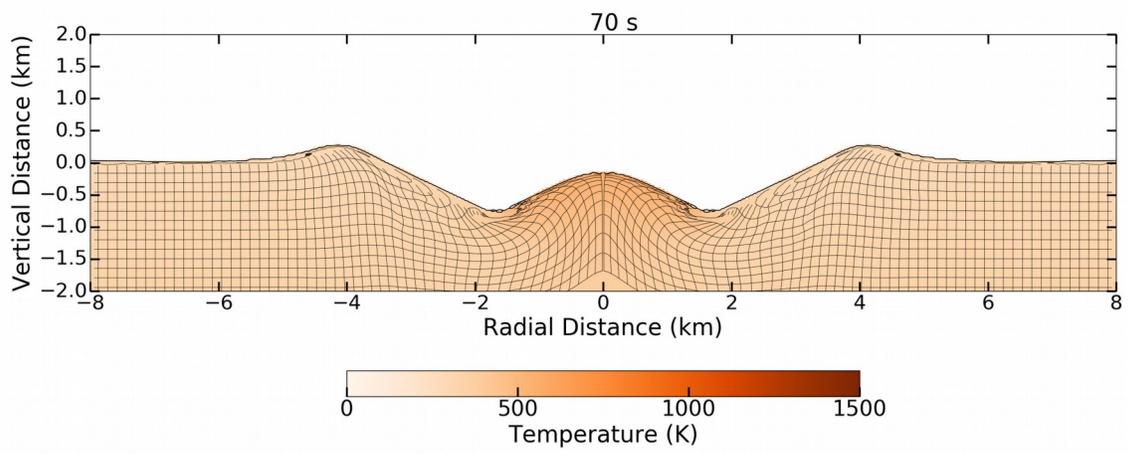

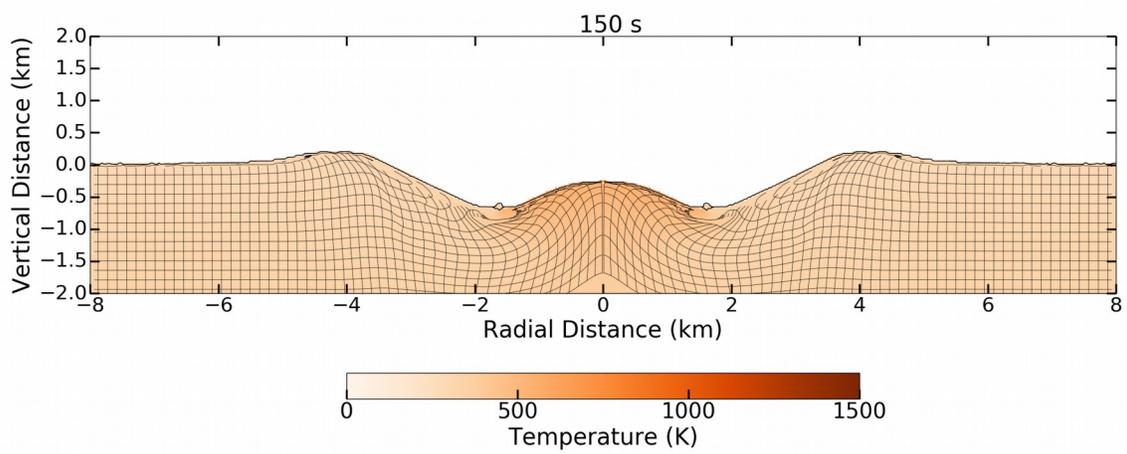

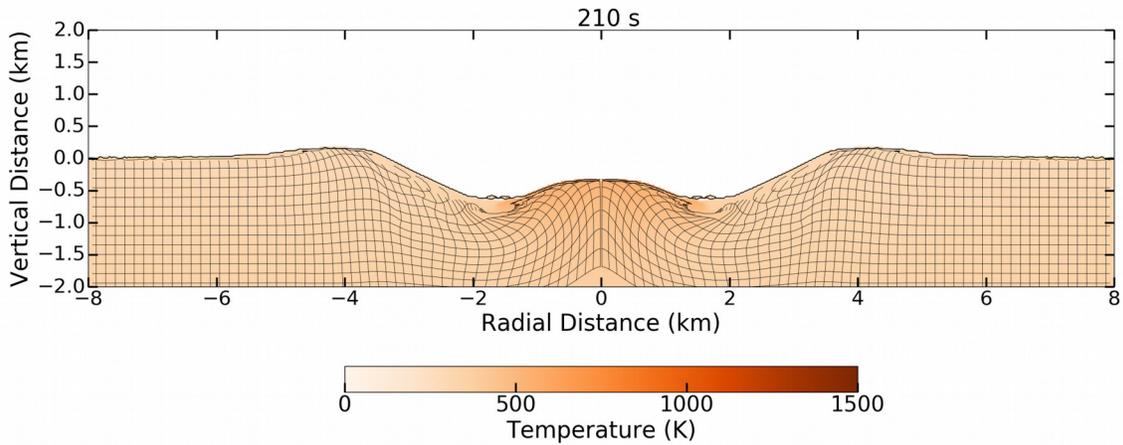

**Figure 8.** The time series of the impact crater as a result of a Centaur from R1 colliding with the Earth. The initial time at 0s is when the collision with the surface starts. It takes ~3 minutes for the crater collapse to finish. The impactor diameter is 1 km. Considering that the simulations are axisymmetric, the impact velocity was scaled to the vertical component only. The non-scaled impact velocity is $v_i$ = 9.4 km s$^{-1}$ (at 45° and scaled to the vertical component is $v_i$ = 6.6 km s$^{-1}$.

However, this amount of water is still significant compared to the amount of with the water in the atmosphere, which is about[15] $8.54*10^{-9}$ the total amount of water on the Earth. We provide a summary of the craters (table 10), and include 3 sets of speeds (average, maximum and minimum) for each impactor size. While it would be speculative to place any constraints on possible Earth craters (based on the confirmed impact structures record) that might have been formed by Centaurs, such possibility should not be entirely dismissed. A large range in crater sizes could be a consequence of very high impact velocities of the Centaurs to the planets, i.e. they arrive to the Earth with an average speed of ~40 km s$^{-1}$, which is at least 1.5 times the average typical velocity of a V-type NEA asteroid impacting the Earth, $v_{imp}$ = 20.0 km s$^{-1}$ (Galiazzo, Wiegert & Aljbaae, 2016).

**Table 10.** Crater dimensions ($D_C$) for each terrestrial planet in kilometers, impactor dimension ($D_i$) in kilometers and impact velocity ($v_i$) in kilometers per second at an impact angle of 45° (minimum, average, maximum).

| Region<br>Planet | $D_{i,min}$ $v_{i,min}$<br>$D_{C,min}$ $v_{i,min}$ | $\bar{D}_i$ $\bar{v}_i$<br>$\bar{D}_C$ $\bar{v}_i$ | $D_{i,max}$ $v_{i,max}$<br>$D_{C,max}$ $v_{i,max}$ | $D_{i,D_{1km}}$ $\bar{v}_i$<br>$D_{C,1km}$ $\bar{v}_i$ ($D_{C,1km,min} - D_{C,1km,max}$) |
|---|---|---|---|---|
| Region 1 | 1.6 | 51.2 | 300.2 | 1.0 |
| Venus | 12.6 12.4 | 484.9 48.9 | 3225.6 84.8 | 15.1 48.9 (8.1-21.1) |
| Earth | 10.5 9.3 | 418.5 38.8 | 2844.1 71.7 | 13.0 38.8 (6.8-18.6) |
| Mars | 10.7 7.2 | 436.7 30.7 | 3005.2 59.1 | 14 30.7 (7.6-19.7) |
| Region 2 | 0.5 | 12.1 | 72.0 | 1.0 |
| Venus | 1.8 11.2 | 133.1 46.3 | 912.9 84.2 | 14.8 46.3 (4.2-10.7) |
| Earth | 3.9 9.8 | 118.7 40.8 | 610.7 71.5 | 13.2 40.8 (7.7-21.0) |
| Mars | 3.0 5.9 | 118.8 27.6 | 862.5 60.2 | 13.5 27.6 (7.0-19.9) |

---

15  1% of the total mass of the atmosphere of the Earth: $5.1*10^{18}$ kg (https://nssdc.gsfc.nasa.gov/planetary/factsheet/earthfact.html).

# 5 CONCLUSIONS

In this paper we presented our results from modelling the total Centaur population to investigate the threat of impacts with terrestrial planets. The Centaur population analysed in our work was subdivided into the biased population (BP), the known observed Centaurs, and the unbiased population (UP), the estimated real population of Centaurs larger than 1 km in size. In particular, 2 regions of Centaurs are emphasized and divided by the mean motion resonance J4:1, at 13.11 au. Region 2 is the source of Centaurs that undergo more encounters (consequently more impacts) with terrestrial planets, particularly in these orbits: 5.5 au ≤ a ≤ 8 au and with eccentricity, 0.09 ≤ e ≤ 0.75 and, i ~ 40°. A minor, but relevant contribution, is in region 1, further divided into 2 sub-regions: (a) a ≤ 20 au, e ≤ 0.63 and I < 35° and, (b) 26 au ≤ a < 28 au and with eccentricity, 0.8 ≤ e ≤ 1 and, 15° ≤ i ≤30°. There is also a considerable contribution to TPEs from Centaurs with retrograde orbits, i ~ 140° and a ~ 22 au. However, the dominant inclination for Centaur sources of TPEs is i ~20°, similar to Jupiter comets. Centaurs can become Jupiter comets in only ~ 20 Kyr. In general, the lifetime of Centaurs is 6.0 Myr. This period of time is comparable to the replenishment of NEOs by the main belt asteroids via the strongest mean motion resonances ($v_6$ and J3:1). This shows that the orbital and physical distribution of the NEOs is strictly dependent on the Centaurs and MBAs together. Thus, distinguishing their orbital evolution can help to understand and characterize different groups of the NEOs and their interaction with other bodies in the solar system. The lifetime for Centaurs in Region 2 is substantially shorter than that of Centaurs in Region 1; these lifetimes for R2 and R1, are 1.8 Myr and 10.3 Myr, respectively. This work shows that in order to study the orbital evolution of the Centaurs, it is not enough to study only one Centaur with its relative clones, as it was performed in previous studies (i.e. Napier, 2015), even if such studies could provide some constraints on the orbital evolution. Centaurs like Chiron can be used as a proxy for its own relative region; in particular, Chiron can be used as a proxy for R1. The majority of the Centaurs escape (82.4%), 6.2% collide with the Sun (of which 56.4% are from Region 2) and 0.9% undergo impacts with planets. For the latter value, Region 2 dominates in a ratio 13:1 over Region 1. 53% of the Centaurs enter the terrestrial planet region, and of these, 57% come from Region 2. They arrive on a timescale of millions years, comparable to the one relative to the main belt bodies in becoming NEOs, which are a mixture of the main belt objects and Centaurs. They survive as NEOs 96 kyr and their contribution to this population is ~4%. This study also confirms that the preferable unbiased population size of km-size Centaurs is 44300, found by Horner, Evans & Bailey (2004a), but not consistent with the higher value found by Sheppard (2000). Our results indicate that the contribution of the Centaurs to the PHOs is about 9%, and for the SPC, it is at least ~31%, with an average dynamical life time (as an SPC) ranging from 0.4 for R1 to 0.6 (R2). Close encounters occur with all the planets, mainly with Jupiter. Among the terrestrial planets, Venus undergoes the most encounters. For the biased Centaurs population from Region 1, about 1 Centaur encounters the TPs every million years, as opposed to Region 2, where 28, 23 and 18 encounters take place every 1 million years, meaning 1 encounter every 36 kyr, 43 kyr and 56 kyr, respectively for Venus, the Earth and Mars. For the unbiased Centaurs population, there is one encounter at a more frequent scale, every 0.7 kyr, 0.8 kyr and 1.0 kyr for Venus, the Earth and Mars, respectively. About 7-8% of all the Centaurs

(the unbiased population) could encounter TPs (the Earth 7.5%). For large bodies (D > 50 km), encounters occur at a rate of about 1 every 40 kyr, and because there are about 11% active Centaurs (11% of all Centaurs become comets), it means tha the probability to have comets arriving in the TP region every 40 kyr is 11%: therefore, there could be comets with a nucleus larger than 50 km, roughly 1 every ~0.4 Myr. For all the Centaurs, we find that in a time frame of 25 Myr, there are 2 time intervals with a high rate of close encounters. The intervals between the two maxima are 8 Myr and 17 Myr. The rate of arrivals for Centaurs coming from Region 2 is 19 times greater than that from Region 1. We find that PHOs with a ≥ 2.6 au and e > 0.63 are potentially former Centaurs, in particular those with a > $a_J$ which should come from Region 1. One example is the asteroid 1999 $XS_{35}$, which could be a former nucleus of a comet or a part of it (diameter = D > 1 km). In addition, comet C/1983 $H_1$ (IRAS-Araki-Alcock) is likely a former Centaur. Centaurs which undergo CE with the Earth have 3 sets of main orbits, with peculiar orbits that mostly reside in the main belt: (i) a ~ 2.7 au, e ~0.64 and i ~9˚, i.e. real small asteroids in range (D < 1000 m) are 2016 $UF_{101}$, 2008 VE, etc. (ii) a ~ 3.2 au, e ~ 0.76 and i ~ 15˚ . Here we find real asteroids and comets in this range, like i.e. 2012 $DG_{61}$ and 222 P/LINEAR. (iii) a ~ 3.4 au, e ~ 0.77 and i ~ 18˚ , e.g. like 2012 $GS_5$ and 74P/Schwassmann-Wachmann group. The Centaurs in close encounters with Venus and Mars orbit, are mainly in the outer main belt at high eccentricities (e ≥ 0.8 for Venus and e ≥ 0.4 for Mars) and i ~ 10˚ and even retrograde orbits. Usually, the Centaurs coming from Region 1 are on the edge of the a-e phase space for the Earth-crossing bodies. The sub-region with inclinations, 85˚ ≤ i ≤ 110 ˚, and a > 11 au is unlikely a source of terrestrial planetary encounters. Objects with a > 20 au and i ≤ 65˚ do not come from Centaurs. Concerning impacts, the impact rate for Centaurs complete the gap found in the main belt objects with sizes larger than ~3 km. Using the Öpik relation, we find that the largest Centaur that can be destroyed completely after activity as a comet is 200 m in diameter. Thus, this result is irrelevant for the rates of impacts computed for the biased (no minor body of the sample is less than 400 m, and the minimum size is ~ 0.47 km) and unbiased populations (D > 1 km). Our results suggest that Region 1 is not a source of catastrophic collisions in the recent solar system (after the LHB). Instead, Region 2 is the likely source of potentially dangerous objects: for a Centaur with an average size of 12.6 km, an impact with the Earth and Venus could happen every 1.898 Gyr and 2.116 Gyr, respectively. This means that 1 or 2 extinction events could have happened on the Earth due to the Centaurs (apart from the main belt objects) in the last 3 billion years. These impacts can be as large as ~130 km on Venus and ~120 km on the Earth, considering an average impact velocity of 46.9 km s$^{-1}$ and 40.9 km s$^{-1}$ , respectively. Although it is not possible to make any firm conclusions, we not that some Earth craters fit these size ranges and are consistent with the timing, i.e., the Sudbury Crater (age 1.85 Gyr and 130 km in diameter) and the Vredefort crater (age 2.023 Gyr and D = 120 km). If these craters had been created by a former Centaur, it follows that we might be presently living in a time epoch when a catastrophic impact event capable of extinction might be possible. For Centaurs larger than 1 km, the unbiased population, there should be on average one impact every ~9.7 Myr, ~8.8 Myr and ~31.0 Myr, for Venus, the Earth and Mars, respectively since the LHB. In the recent solar system, however, it should happen less frequently, approximately every 14.4 Myr, 13.1 Myr and 46.3, respectively for Venus, the Earth and Mars. On the Earth, the craters should be at least 7 km in diameter for these kind of impacts.

In terms of water release to the Earth, we compute that Centaurs could release a value of roughly $10^{-5}$ of the total water present now since the LHB and, at the moment they should release to our planet $2*10^{-10}$ (current) Earth-water-like content every million year and for Mars at least less than half of this value. It is interesting to note that this result is also applicable to Mars, where the water delivery is $3*10^{-5}$ the total water present on the Earth now, the amount consistent with the amount for a "dry" planet (and only in the recent solar system).

Our work suggests that the Centaurs are very important for the evolution of a planet, and could have profound implications geologically and also astrobiologically, i.e. we can not esclude that the K/T event was caused by a Centaur. Finally, we would like to invite the scientific community to observe more this population of objects given that it is of vital impact to our planet (and also to Mars), because we need the tools to better discriminate their size distribution (and thus a better estimate of their impact rate) and their physical characteristics, the amount of water and other important material they might bring to the terrestrial planets.

## ACKNOWLEDGEMENTS

MG wishes to acknowledge the support by the Austrian FWF project J-3588-N27, and thanks Y. Cavecchi for suggestions to improve the computations; S. Aljbaae running some of our numerical orbital integrations on a cluster at FEGP-UNESP (Brazil); and Western University for providing computational facilities (the TITAN supercluster and the SHARCNET). EAS gratefully acknowledges the Natural Sciences and Engineering Research Council of Canada (NSERC) Postdoctoral Fellowship program for supporting this project. RD acknowledges the support by the Austrian FWF project S11603-N16. We gratefully acknowledge the developers of iSALE-2D (www.isale-code.de), the simulation code used in our research, including G. Collins, K. Wunnemann, D. Elbeshausen, B. Ivanov and J. Melosh. The authors also thank the referee Mario Melita for valuable advice allowing us to improve this paper.

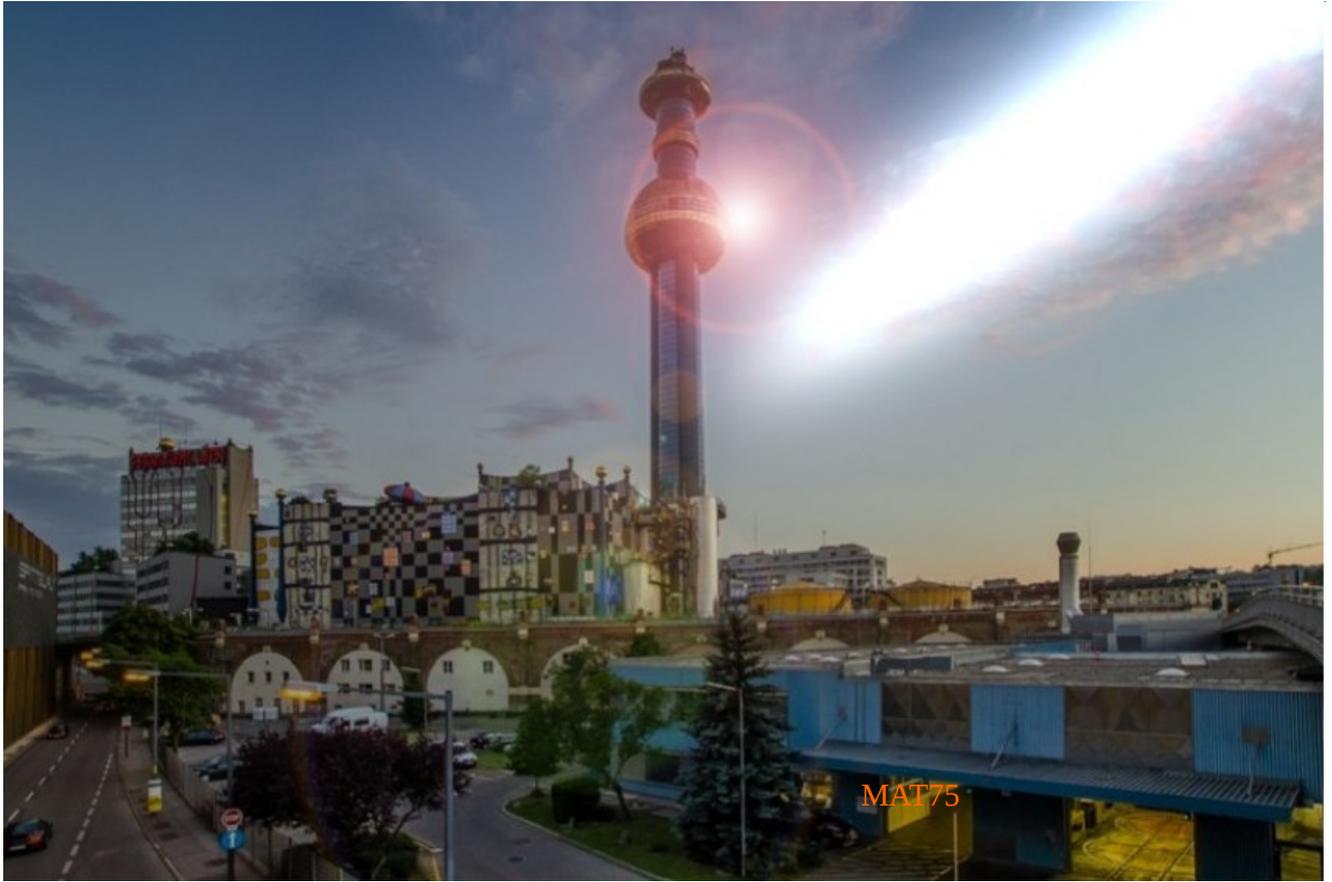